\documentclass[aps,superscriptaddress,eqsecnum,nofootinbib,showpacs,preprintnumbers]{revtex4-2}
\usepackage[utf8]{inputenc}
\usepackage{amsmath}
\usepackage{amssymb}
\usepackage{amsfonts,amsthm,bm}
\usepackage{color,xcolor}
\usepackage{comment}
\usepackage{soul}

\usepackage{graphicx,epsfig}
\usepackage{float}
\usepackage{subcaption}
\usepackage{tikz}
\newcommand{\be}{\begin{eqnarray}}
	\newcommand{\ee}{\end{eqnarray}}
\newcommand{\bea}{\begin{eqnarray}}
	
	\newcommand{\eea}{\end{eqnarray}}

\usepackage{ulem}

\def\Ref{\ref}

\definecolor{azure(colorwheel)}{rgb}{0.0, 0.5, 1.0}
\definecolor{DarkViolet}{RGB}{148,0,211}
\definecolor{myDarkBlue}{rgb}{0,0.1,0.7}
\definecolor{DarkBlue}{RGB}{0,0,153}
\definecolor{amber}{rgb}{1.0, 0.49, 0.0}
\definecolor{amaranth}{rgb}{0.9, 0.17, 0.31}
\definecolor{nicered}{rgb}{0.7,0.1,0.1}
\definecolor{brown}{rgb}{0.5,0.1,0.1}
\definecolor{nicegreen}{rgb}{0.0,0.3,0.0}
\definecolor{tealgreen}{rgb}{0.0, 0.51, 0.5}

\definecolor{tclr}{RGB}{148,0,211}

\usepackage{hyperref}
\hypersetup{colorlinks,bookmarksopen,
	bookmarksnumbered,
	citecolor={nicered},
	linkcolor={myDarkBlue},
	urlcolor={tealgreen},
	pdfstartview=FitH}

\newcommand{\beq}{\begin{equation}}
\newcommand{\eeq}{\end{equation}}
\newcommand{\bseq}{\begin{subequations}}
	\newcommand{\eseq}{\end{subequations}}

\usepackage{times}

\graphicspath{{Figs/}}

\usepackage{orcidlink}
\def\idkara{\orcidlink{0000-0002-5479-6513}}
\def\idbako{\orcidlink{0000-0002-3012-6144}}

\begin{document}
\title{Thermodynamic analysis of shift-symmetric black-hole spacetimes in Horndeski gravity}

\author{Athanasios Bakopoulos\idbako}
	\email{atbakopoulos@gmail.com}
    \affiliation{Division of Applied Analysis, Department of Mathematics, University of Patras, Rio Patras GR-26504, Greece}
	\affiliation{Physics Department, School of Applied mathematical and Physical Sciences,
	National Technical University of Athens, 15780 Zografou Campus,
	Athens, Greece.}

\author{Thanasis Karakasis\idkara}
	\email{thanasiskarakasis@mail.ntua.gr}
	\affiliation{Physics Department, School of Applied mathematical and Physical Sciences,
	National Technical University of Athens, 15780 Zografou Campus,
	Athens, Greece.}

\begin{abstract}
    In this study, we investigate the thermodynamic properties of shift-symmetric Horndeski and beyond Horndeski theories (theories in which only derivatives of the scalar field appear in the action). Utilizing Euclidean methods, we first analyze two specific cases that serve as foundational examples for the broader framework. We derive the general expression for entropy variation within this setting, for homogeneous spacetimes (with $g_{tt}=1/g_{rr}$), demonstrating that both parity-preserving and parity-violating terms contribute to the entropy formula. Given the functional form of the coupling terms, the black hole entropy can be determined by integrating this relation with respect to the event horizon. Our general result is consistent with previously reported special cases in the literature. Notably, for particular constraints on the coupling functions, the entropy identically vanishes. Furthermore, we establish that homogeneous spacetimes within shift- and parity-symmetric beyond-Horndeski theories universally obey Bekenstein’s Area Law for entropy, irrespective of the specific form of the coupling functions.
\end{abstract}

\maketitle

\section{Introduction}

Scalar-tensor theories provide a natural extension of General Relativity (GR), incorporating a scalar degree of freedom that influences gravitational dynamics. Among them, Horndeski theory \cite{Horndeski:1974wa} represents the most general four-dimensional scalar-tensor theory leading to second-order equations of motion for both the metric and the scalar field, ensuring the absence of ghost-like instabilities. Originally formulated in the 1970s, Horndeski gravity has since gained significant attention as a unifying framework for modified theories of gravity. Its generality allows for a broad spectrum of gravitational phenomena, accommodating deviations from GR while maintaining theoretical consistency. A fundamental result in the study of scalar-tensor theories is that every well-defined extension of GR that introduces a single scalar field can be mapped to a subclass of Horndeski theory, reinforcing its fundamental role in gravitational physics. Consequently, any viable scalar-tensor model—whether proposed to explain inflation, cosmic acceleration, or strong-field modifications to gravity—inevitably finds a description within the Horndeski framework.

Several well-known theories of modified gravity emerge as specific cases of Horndeski gravity, illustrating its broad applicability. For instance, the covariant Galileon  model \cite{Nicolis:2008in, Deffayet:2009mn, Deffayet:2009wt, Deffayet:2011gz}, which introduces a shift symmetry in the scalar field, is naturally contained within Horndeski theory and has been extensively studied in the context of early universe dynamics. Similarly, Einstein-dilaton-Gauss-Bonnet gravity \cite{Kanti:1995vq}, which arises in string theory-inspired modifications of GR, belongs to the Horndeski class and has been explored in the context of black hole physics and compact objects \cite{Antoniou:2017acq,Doneva:2017bvd,Silva:2017uqg,Antoniou:2017hxj}. Other notable examples include Kinetic Gravity Braiding models \cite{Deffayet:2010qz,Pujolas:2011he}, which introduce non-trivial interactions between the kinetic term of the scalar field and gravity, leading to distinctive cosmological \cite{Kobayashi:2010cm,Appleby:2011aa,Creminelli:2008wc} and astrophysical signatures \cite{Babichev:2013cya, Babichev:2016rlq, Minamitsuji:2016hkk}. Due to this versatility, Horndeski theory has played a central role in the study of gravitational physics, offering a rich theoretical landscape for investigating both fundamental and observational aspects of gravity.

A particularly important subclass of Horndeski gravity is shift-symmetric Horndeski theory, where the scalar field enters the action only through its derivatives. This additional symmetry has profound theoretical and practical implications. Many shift-symmetric scalar-tensor models naturally arise from higher-dimensional theories, such as compactifications of Lovelock gravity \cite{Charmousis:2012dw,Lecoeur:2024kwe,Lu:2020iav,Glavan:2019inb,Fernandes:2021dsb,Alkac:2022fuc}, where the shift symmetry emerges as a remnant of higher-dimensional diffeomorphism invariance. Similarly, low-energy limits of Einstein-dilaton-Gauss-Bonnet gravity often exhibit shift symmetry \cite{Sotiriou:2013qea,Sotiriou:2014pfa}, reinforcing its significance in effective field theory descriptions of modified gravity. Beyond its theoretical motivation, shift symmetry plays a crucial role in simplifying the field equations, often enabling the derivation of exact analytical solutions that would otherwise be inaccessible \cite{Bakopoulos:2022csr}. This has been particularly important for the discovery of stealth black hole solutions, where the scalar field has a non-trivial configuration while leaving the metric identical to GR solutions \cite{Babichev:2013cya,Minamitsuji:2019shy}. Furthermore, shift symmetry is essential for constructing solutions with primary scalar hair \cite{Bakopoulos:2023fmv,Baake:2023zsq,Bakopoulos:2023sdm}, providing a mechanism through which black holes can support an independent scalar charge, while, besides its application in black-hole physics it has also been used in order to alleviate cosmological tensions \cite{Petronikolou:2021shp}. 

A significant motivation for extending Horndeski gravity arises from the so-called beyond-Horndeski theories. These models introduce additional higher-order interactions that, while formally extending the Horndeski action, still evade Ostrogradsky instabilities and preserve the well-posedness of the equations of motion \cite{Gleyzes:2014dya,BenAchour:2016cay}. The necessity of such extensions became evident when it was realized that certain degenerate higher-order scalar-tensor theories could maintain a second-order effective structure, despite their apparent inclusion of higher derivatives. These modifications allow for new phenomenology in gravitational physics, particularly in the strong-field regime, where deviations from GR are expected to be most prominent \cite{Kobayashi:2019hrl}. Beyond-Horndeski theories also broaden the scope of possible solutions in scalar-tensor gravity, allowing for novel black hole geometries, exotic compact objects, and modifications to gravitational collapse. As a result, they have become a key area of study in modern theoretical gravity, providing insights into both astrophysical and cosmological phenomena \cite{Babichev:2016rlq,Gleyzes:2015pma,Creminelli:2017sry}.

Horndeski theory has been extensively utilized in the study of exact gravitational solutions, leading to the discovery of black holes, wormholes, and solitonic configurations that deviate from their counterparts in GR. Black hole solutions in this framework have been of particular interest, as they offer testable deviations from the Kerr \cite{Charmousis:2019vnf,Anson:2020trg,Walia:2021emv,Maselli:2015tta,Kleihaus:2011tg} and Schwarzschild metrics \cite{Bakopoulos:2020dfg,Bakopoulos:2018nui,Bakopoulos:2023hkh,Bakopoulos:2021dry,Karakasis:2021lnq,Karakasis:2021rpn,Karakasis:2021ttn,Karakasis:2023hni,Karakasis:2023ljt,Karakasis:2021ttn,Tang:2020sjs,Kiorpelidi:2022kuo,Ventagli:2020rnx,Thaalba:2022bnt,Antoniou:2022agj,Antoniou:2021zoy,Antoniou:2020nax,Andreou:2019ikc} potentially revealing new gravitational effects in the strong-field regime. Traditionally, black holes in scalar-tensor theories were thought to obey strict no-hair theorems, which prohibit the existence of independent scalar charges. However, recent studies have demonstrated that within Horndeski gravity, black hole solutions featuring primary scalar hair can indeed exist \cite{Bakopoulos:2023fmv,Baake:2023zsq,Bakopoulos:2023sdm,Guajardo:2024hrl,Charmousis:2025jpx}. These configurations are particularly intriguing, as the scalar field is not merely a secondary effect of the black hole’s mass or charge but instead constitutes an independent characteristic of the solution. Such discoveries challenge the traditional understanding of black hole structure and open new avenues for exploring deviations from GR. In particular, in the same shift-symmetric theories that support the primary hair, a large number of novel regular solutions, i.e. solutions without an initial singularity, have been found \cite{Bakopoulos:2023fmv,Bakopoulos:2023sdm}. In addition to black holes, Horndeski gravity has also been employed in constructing wormhole solutions, where the scalar field plays a crucial role in supporting the exotic geometry \cite{Bakopoulos:2022csr,Bakopoulos:2021liw,Bakopoulos:2023tso,Antoniou:2019awm,Kanti:2011jz,Kanti:2011yv,Karakasis:2021tqx,Chew:2016epf,Chew:2018vjp,Blazquez-Salcedo:2018ipc,Chew:2019lsa,Chew:2020lkj,Blazquez-Salcedo:2020nsa}, and solitonic solutions, which describe localized, self-stabilizing field configurations in curved spacetime \cite{Bakopoulos:2022csr,Bakopoulos:2023fmv,Bakopoulos:2023sdm,Kleihaus:2019rbg,Kleihaus:2020qwo}. The emergence of these solutions highlights the richness of Horndeski theory and its potential to describe novel gravitational phenomena beyond standard GR predictions.

The study of black hole thermodynamics has been one of the most profound developments in theoretical physics, revealing deep connections between gravity, quantum mechanics, and statistical physics. The pioneering works of Bekenstein \cite{Bekenstein:1972tm, Bekenstein:1973ur} and Hawking \cite{Hawking:1976de,Hawking:1975vcx} established that black holes are not merely gravitational ``sinks’’, but thermodynamic objects characterized by temperature and entropy. The identification of black hole entropy with the horizon area and the formulation of the four laws of black hole mechanics \cite{Bardeen:1973gs} provided a thermodynamic interpretation of gravitational dynamics, suggesting a deeper microscopic origin for gravitational entropy. This realization has been instrumental in understanding the interplay between quantum field theory and gravity, ultimately motivating attempts to formulate a consistent theory of quantum gravity \cite{Carlip:2014pma}.

Black hole thermodynamics is essential for probing the stability, phase transitions, and information content of black hole solutions. Extracting the thermodynamic properties of black holes allows us to determine their response to perturbations, their ability to exchange energy with their environment, and further probe the role of gravity at the quantum realm, even at a ``semi-classical" level.
 In the context of modified gravity, studying black hole thermodynamics can reveal whether alternative theories respect fundamental principles such as the area law of entropy, the first law of thermodynamics, or under what conditions modified gravity manifests at the thermodynamic level. For example, possible modifications on the entropy may have implications for holography and quantum gravity, and introducing puzzles in the theoretical physics community such as the black hole information paradox \cite{Giddings:1995gd, Mathur:2008wi}.

In scalar-tensor theories, and particularly in Horndeski gravity, black hole thermodynamics plays a fundamental role in understanding the interplay between additional scalar degrees of freedom and gravitational dynamics. Since scalar-tensor theories introduce non-minimal couplings between the scalar field and curvature, they generally modify the standard thermodynamic properties of black holes. The entropy, temperature, and free energy can acquire novel corrections, leading to distinct thermodynamic behavior compared to GR solutions. A characteristic example concerns Bekenstein's area law for black hole entropy \cite{Bekenstein:1972tm,Bekenstein:1973ur}, which states that the entropy of a black hole is proportional to the area of the event horizon. This observation was made possible due to Hawking's proof that the area of the event horizon never decreases \cite{Hawking:1971tu}. In particular, it has been showed so far that the entropy of black holes in scalar-tensor theories may not follow the area law and instead obey a modified area law due to the fact that Newton's constant has been promoted to a field that depends on the scalar degree of freedom introduced by modified gravity \cite{Brustein:2007jj}. Particular examples of this behavior include black holes with conformally coupled scalar fields \cite{Martinez:1996gn,Karakasis:2021rpn,Karakasis:2021ttn,Martinez:2005di,Barlow:2005yd,Anastasiou:2022wjq, Bravo-Gaete:2025vyd}, black holes in $f(R)$ gravity \cite{Sebastiani:2010kv, Faraoni:2010yi} and Brans-Dicke gravity \cite{Cai:1996pj}, a phenomenon which however is absent in the case of minimally coupled scalar fields \cite{Chew:2022enh, Chew:2023olq, Chew:2024rin, Chew:2024evh, Karakasis:2023hni, Karakasis:2022fep, Gonzalez:2013aca, Bakopoulos:2021dry, Bakopoulos:2023hkh, Karakasis:2022xzm}. Such behavior has been observed in general in modified gravity \cite{Cai:2001dz, Jacobson:1993vj, Tachikawa:2006sz}. Investigating these modifications is crucial for determining whether Horndeski gravity respects established thermodynamic principles. 

The thermodynamics of black holes in Horndeski gravity has been an active area of research, with several works addressing this topic from different perspectives. Early studies focused on specific black hole solutions, extracting their entropy and temperature within particular subclasses of Horndeski theory \cite{Feng:2015oea, Bravo-Gaete:2014haa, Bravo-Gaete:2021hlc, Bravo-Gaete:2021hza}. The thermodynamics of regular black holes arising from an infinite tower of curvature corrections was studied more recently in \cite{Cisterna:2025vxk} (for interesting studies regarding this matter and regular black holes arising in the context of pure gravity theories we refer to \cite{Hennigar:2025ftm, Bueno:2025zaj, Bueno:2024dgm, Fernandes:2025eoc, Fernandes:2025fnz}). Efforts have also been made to understand black hole thermodynamics in the most general setting possible. Notable examples include works such as \cite{Minamitsuji:2023nvh}, which explored the thermodynamic properties of black holes in Horndeski gravity using the Iyer-Wald formalism \cite{Wald:1993nt, Iyer:1994ys} providing a detailed study of black hole thermodynamics in general Horndeski theories. However, given the vast complexity of Horndeski and beyond-Horndeski theories, further investigation is needed to develop a systematic method for extracting thermodynamic properties in a broad class of solutions.

In this work, we develop a general approach to extract black hole thermodynamics in shift-symmetric Horndeski and beyond-Horndeski theories by applying the Euclidean method. Our method allows us to compute the thermodynamic quantities directly from the action, providing a systematic framework for analyzing black hole solutions in these extended theories. By implementing this formalism, we derive the entropy, the temperature, and the mass for a wide class of black hole spacetimes, including those with non-trivial scalar hair. This approach not only unifies existing thermodynamic analyses but also provides new insights into the role of shift symmetry and beyond-Horndeski modifications in black hole physics. Our results contribute to the broader effort of understanding black hole thermodynamics in modified gravity and serve as a stepping stone for future studies on the thermodynamic stability and phase transitions of black holes in scalar-tensor theories. 
In particular, we have derived a general expression for the variation of the entropy in any shift-symmetric beyond-Horndeski theory. The black hole entropy can be determined by simply integrating this relation with respect to the event horizon radius, ensuring that it vanishes in the absence of the black hole (see section \ref{sec4}) This general result, applies to any homogeneous metric and has been tested by us in various well-known black-hole solutions. As is already pointed out in \cite{Jacobson:1993vj, Feng:2015oea, Bakopoulos:2024zke, Minamitsuji:2023nvh}, the Wald entropy formula \cite{Wald:1993nt} may not be applicable in the Horndeski framework, due to the covariant derivatives of the scalar field in the action, while the Wald method may lead to inconsistent results \cite{Feng:2015oea}. The Wald entropy formula has been tested many times in the literature so far and in general yields results that agree with Euclidean methods. However, as is pointed out in \cite{Hajian:2020dcq, Jacobson:1993vj} in several theories an inconsistency may arise. In particular, Ref.~\cite{Jacobson:1993vj} showed that the ambiguity is harmless for stationary black holes with a regular Killing horizon, but in scalar–tensor theories the regularity assumption can fail, e.g. due to divergences of scalar derivatives at the horizon \cite{Hajian:2020dcq}, and the Wald formula may then become ill-defined. The theories are equivalent (yield the same equations of motion) up to boundary terms and the Wald formula relies on the derivative of the Lagrangian with respect to the Riemann tensor, while treating the metric as an independent quantity. It is not clear whether taking this derivative of a theory, before or after integration by parts will give the same result for the Wald entropy and this seems to be the reason for inconsistent results. The Euclidean method evades these problems, as is also discussed in \cite{Bakopoulos:2024zke}. The sole requirement of the Euclidean procedure is the attainment of a well-defined variational procedure by adding the appropriate boundary term and our approach relies on the fact that we obtain the exact form of the boundary term that needs to be added in order to achive this. 

This work is structured as follows. In Section \ref{sec2}, we introduce the class of shift- and parity-symmetric scalar-tensor theories under consideration and discuss their theoretical motivation. Section \ref{thermoexample} focuses on specific black hole solutions within these theories, emphasizing the role of shift and parity symmetry in their existence and thermodynamic properties. In Section \ref{sec4}, we develop a systematic approach for extracting thermodynamic properties using the Euclidean method, applying it to black hole solutions in both Horndeski and beyond-Horndeski theories. Section \ref{sec5} presents a detailed discussion of our results, comparing the entropy structure in different cases and highlighting the implications of parity breaking. Finally, we summarize our conclusions and discuss potential future directions in Section \ref{conc}.

\section{Theoretical Framework}\label{sec2}

In what follows we will consider shift symmetric Horndeski plus beyond Horndeski theory. The former  is parametrized by four functions $\{G_i: i=2,..,5\}=\{G_2,G_3,G_4,G_5\}$ of  $\phi$ and its kinetic energy density $X=-\frac{1}{2}\partial_\mu\phi\,\partial^\mu\phi$ :
\be
\label{eq:Hfr}
S_{H} = \displaystyle\int d^4x \sqrt{-g} \,\left(\mathcal{L}_2+\mathcal{L}_3+\mathcal{L}_4+\mathcal{L}_5\right),
\ee
with
\begin{align}
\mathcal{L}_2 &=G_2(X) ,
\label{eq:L2fr}
\\
\mathcal{L}_3 &=-G_3(X) \,\Box \phi ,
\label{eq:L3fr}
\\
\mathcal{L}_4 &= G_4(X)R+G_{4X}[(\square\phi)^2-\phi_{\mu\nu}\phi^{\mu\nu}] ,
\label{eq:L4fr}
\\
\mathcal{L}_5 &= G_5(X) G_{\mu\nu}\phi^{\mu\nu} - \frac{1}{6}\, G_{5X} \big[ (\Box \phi)^3 - 3\,\Box \phi\, \phi_{\mu\nu}\phi^{\mu\nu}+ 2\,\phi_{\mu\nu}\phi^{\nu\rho} \phi_{\rho}^{~\mu} \big].
\label{eq:L5fr}
\end{align}
The latter is given by two additional higher order terms,
\begin{align}
\mathcal{L}^{\rm bH}_4&=F_4(X)\varepsilon^{\mu\nu\rho\sigma}\,\varepsilon^{\alpha\beta\gamma}_{\,\,\,\,\,\,\,\,\,\,\,\sigma}\,\phi_\mu\,\phi_\alpha\,\phi_{\nu\beta}\,\phi_{\rho\gamma},\\[3mm]
\mathcal{L}^{\rm bH}_5&=F_5(X)\varepsilon^{\mu\nu\rho\sigma}\,\varepsilon^{\alpha\beta\gamma\delta}\,\phi_\mu\,\phi_\alpha\,\phi_{\nu\beta}\,\phi_{\rho\gamma}\,\phi_{\sigma\delta},
\end{align}
forming the action $S_{\rm bH}=\displaystyle\int d^4x \sqrt{-g} \,\left(\mathcal{L}^{\rm bH}_4 + \mathcal{L}^{\rm bH}_5\right)$, hence the total action of our work will be given by
\begin{equation}
    S = S_H + S_{\rm bH}~. \label{totalaction}
\end{equation}
Here, the notation is defined as follows: $\phi_{\mu} = \partial_{\mu}\phi,~\phi_{\mu\nu} = \nabla_{\mu}\partial_{\nu}\phi$ and subscripts denote derivative with respect to the argument: $~ G_{iX} = \partial_X G_i$. GR is included in the $\mathcal{L}_4$ term, when $G_4$ contains a constant, $X-$independent term.
The beyond Horndeski terms parametrized by $F_4$ and $F_5$ are not independent. They are related so as to evade the appearance of a ghost degree of freedom \cite{Crisostomi:2016tcp}. 
This relation reads
\begin{equation}
    X G_{5X} F_4= 3 F_5 (G_4-2X G_{4X})\,. \label{noghost}
\end{equation}
We can see that when $G_{5X}=0$ or $F_4=0$, the degeneracy condition Eq. (\ref{noghost}) implies that either $F_5=0$ or $G_4 \sim \sqrt{X}$. So, in this case, a non-vanishing $F_5$ fixes the form of the coupling function $G_4$ which as we can see does not include GR, since a constant term is absent. Moreover, since we are interested in shift-symmetric theories, our general action will involve only derivatives of the scalar field. As a result, imposing that the scalar field $\phi$ inherits the spacetime symmetries $\phi=\phi(r)$, we may substitute the radial derivative of the scalar field with the kinetic energy $X$, via the very definition of the kinetic energy
\begin{equation}
    \phi'(r) = s \sqrt{-2Xg_{rr}}~, \label{phidot}
\end{equation}
where $s=\pm 1$. In the cases where, besides the shift symmetry we also have parity symmetry, hence an even number of derivatives appear in the action and the theory will be defined by $G_3=0=G_5=F_5$, the value of $s$ will not bother us. However, when this is not the case, the sign of $s$ plays a role and one may choose its sign by imposing a well-defined black hole spacetime (well-defined asymptotics for example). 
Before dealing with the general case and making general arguments, let us set the stage by discussing the thermodynamics of a few particular examples of black hole spacetimes, presented in \cite{Bakopoulos:2022csr}.

\section{Thedmodynamics of Beyond Horndeski theories}\label{thermoexample}


In this section, we provide a detailed discussion of the thermodynamics of black hole solutions in beyond-Horndeski theories, focusing on their fundamental properties and thermodynamic behavior. Our goal is to analyze specific cases that serve as illustrative examples, paving the way for a broader investigation of black hole thermodynamics in these modified gravity theories. By employing Euclidean methods, we examine the key contributions to the black hole entropy, temperature, and free energy, emphasizing the role of shift symmetry.  We will examine two particular cases: one that possesses an additional shift symmetry and another that explicitly breaks this symmetry.

\subsection{Shift and Parity Symmetric Theories}

The action of shift and parity symmetric beyond Horndeski theories will be given by \cite{Bakopoulos:2022csr} 
\begin{equation}
    S= \int d^4x\sqrt{-g}\left[G_4(X)R+G_{4X}[(\square\phi)^2-\phi_{;\mu\nu}\phi^{;\mu\nu}]+G_2(X)+F_4(X)\epsilon^{\mu\nu\rho\sigma}\epsilon^{\alpha\beta\gamma}_{\,\,\,\,\,\,\,\,\,\sigma}\phi_{\mu}\phi_{\alpha}\phi_{\nu\beta}\phi_{\rho\gamma}\right]~. \label{beyondactionparity}
\end{equation}
The theory (\ref{beyondactionparity}) besides containing only derivatives of the scalar field, possessing as a result shift symmetry $\phi\to\phi+c$ with $c$ being a constant, contains also an even number of derivatives for the scalar field, hence it is also characterized by parity symmetry $\phi\to-\phi$. These two symmetries have been exploited very recently in the search for black hole solutions featuring a primary scalar hair \cite{Bakopoulos:2023fmv,Baake:2023zsq,Bakopoulos:2023sdm,Bakopoulos:2024ogt}. Now let us fix the coupling functions in order to discuss the thermodynamics of a particular model, presented in \cite{Bakopoulos:2022csr} for the first time.

\subsubsection{The theory and the solution}

The coupling functions
\begin{equation}
    G_2 = -\varepsilon \mu X^2~,~G_4 = 1-\frac{1}{2} X (\delta  \zeta -\alpha
   )-\frac{1}{2} \delta  \mu  X^2~,~ F_4=\frac{3 \delta  \mu }{8}-\frac{\alpha -\delta  \zeta
   }{8 X}~, \label{functions}
\end{equation}
have been used in \cite{Bakopoulos:2022csr} for the derivation of analytical solutions. We can solve the field equations in order to obtain the spacetime element as well as the form of the scalar field, which we will assume that inherits the spacetime symmetries and therefore is a function of the radial coordinate. Here $\alpha,\delta,\varepsilon,\zeta,\mu$ are constants of the theory. Note, that only derivatives of the scalar field appear in the action and hence we are free to substitute $\phi'(r)$ with $X(r)$, using relation (\ref{phidot}). Assuming the line element 
\begin{equation}
    ds^2 = -f(r)dt^2 + \frac{dr^2}{f(r)} + r^2 d\Omega^2~, \label{lorenttzianmetric}
\end{equation}
the aforementioned action functional is achieved stationary for 
\begin{eqnarray}
    &&f(r) = 1-\frac{2 m}{r}-\frac{\pi  \sqrt{\frac{\varepsilon }{\delta }} (\alpha -\delta  \zeta )^2}{16 \varepsilon  \mu  r}+\frac{(\alpha -\delta  \zeta )^2 \tan ^{-1}\left(\frac{\sqrt{\varepsilon } r}{\sqrt{\delta }}\right)}{8 \sqrt{\delta } \sqrt{\varepsilon } \mu  r}~, \label{f}\\
   &&X(r) = \frac{\alpha -\delta  \zeta }{2 \mu  \left(\delta +\varepsilon  r^2\right)} \label{x}~,
\end{eqnarray}
with $m$ being an integration constant related to the ADM mass of the black hole $\lim_{r\to \infty}(r-rf(r))/2 = m$~. Note that the form of the kinetic energy does not contain the integration constant $m$ and that its behavior is solely determined by the constants of the theory. When $\alpha=\delta\zeta$, $X=0$ and $f(r)=1-2m/r$, hence we go back to the Schwarzschild solution.

\subsubsection{Euclidean Action}

To discuss the thermodynamics of this theory and black hole solution, we utilize the ADM decomposition of a static spacetime with the metric 
\begin{equation}
    ds^2 = N(r)^2 f(r) d\tau^2 + \frac{dr^2}{f(r)} + r^2 d\Omega^2~, \label{adm}
\end{equation}
where we have introduced the lapse function $N$.
The partition function for a thermodynamic system will be given by 
\begin{equation}
    \mathcal{Z} = \int d[g_{\mu\nu}^E,\psi]e^{-\mathcal{I}_E}~,
\end{equation}
where the subscript $E$ denotes Euclidean quantities, and $\psi$ collectively represents the matter fields. The spacetime element has an Euclidean signature $(+,+,+,+)$, and we can transition to this signature by performing a Wick rotation $t \to i\tau$. If we are considering a black hole spacetime, then to avoid the conical singularity at the horizon of the black hole $r_h$, the time coordinate $\tau$ must be periodic, with a period given by \cite{Bakopoulos:2024hah,Karakasis:2023hni}
\begin{equation}
    \beta = \frac{4\pi}{\sqrt{g_{\tau\tau}^{~'}(r) g^{rr~'}(r)}}\Bigg|_{r_h} ~, \label{period}
\end{equation}
which for our metric element becomes $\beta = 4\pi/(N(r)f'(r))|_{r_h}$
so that $0 \leq \tau \leq \beta$. This periodicity will be associated with the inverse temperature of the black hole, and as one can see, the temperature of the black hole solution will be determined entirely from the spacetime metric. Moreover, black holes of different sizes will correspond to different temperatures; however, some exotic cases have also been encountered \cite{Bakopoulos:2024hah}. The free energy of a statistical system, $\mathcal{F}$, will be related to the partition function via \cite{Gibbons:1976ue}
\begin{equation}
    \mathcal{F} = -\ln \mathcal{Z}/\beta~. \label{euclid}
\end{equation}
By imposing the saddle point approximation, we may consider that for the black hole spacetime that will concern us in this work, it is sufficient to consider the contribution of the action of the theory evaluated on the classical solution when the field equations hold: $\delta \mathcal{I}_E = 0$~. Consequently, we need to calculate the on-shell action of our theory, ensuring it attains a true extremum within the class of fields considered. From this, we can relate the Euclidean action to the free energy of the system via
\begin{equation}
    \mathcal{I}_E = \mathcal{F}\beta~. \label{action_free_energy}
\end{equation}

By using the Arnowitt–Deser–Misner (ADM) decomposition, the action (\ref{beyondactionparity}) of the theory takes the following form
\begin{equation}
    \mathcal{I}_E = \int d^3x d\tau \left[ P\dot{\phi} + \Pi^{ij}\dot{g}_{ij} -NH -N^i H_i \right] + \mathcal{B}~, \label{actionE}
\end{equation}
where $P$ is the corresponding conjugate momentum of the scalar field, $\Pi$ is the momentum of the metric, $N$ is the lapse function, $N^i$ is the shift vector, and $H,~ H_i$ are the Hamiltonian densities. $\mathcal{B}$ denotes the boundary terms required to ensure a well-defined variational problem. For a very pedagogical explanation on why the boundary terms should be included, we refer the reader to \cite{Witten:2024upt}. Since both the scalar field and the spacetime metric are time independent, while the spacetime is non-rotating we can consider the reduced action principle 
\begin{equation}
    \mathcal{I}_E = - 4\pi \beta\int_{r_h}^{\infty} dr NH + \mathcal{B}~,
\end{equation}
where the reduced Hamiltonian  $H$ is a quantity that acts as a Hamiltonian constraint and vanishes when the field equations hold (on-shell), while $N$ acts as a Lagrange multiplier and we have performed the integrations over the angles. To compute the before-given action, we substitute the Euclidean line element (\ref{adm}) and obtain the following reduced Hamiltonian, 
\begin{equation}
    H = -2 r f'(r)+2 f(r)-\mu  X(r)^2
   \left(\delta +\varepsilon  r^2\right)-X(r) (\delta
    \zeta -\alpha )+2~.
\end{equation}
This simple form of reduced Hamiltonian can be achieved by exploiting the fact that two theories are equivalent up to boundary terms. Hence when substituting the ADM metric (\ref{adm}) in the Euclidean action (\ref{actionE}) we encounter terms like $\int dr (N''(r) \times A(r))$ where $A(r)$ is a function of $X,f$ and their derivatives. Such a term may be written as $\int dr (A''(r)\times N(r))$ where we have performed integration by parts and canceled the total derivatives that appeared.  Consequently, we may rewrite the action (\ref{actionE}) in terms of the dynamical fields $N,f,X$ as
\begin{equation}
     \mathcal{I}_E = 4\pi\beta \int dr N(r) \left(2 r f'(r)+2 f(r)+\mu  X(r)^2
   \left(\delta +\varepsilon  r^2\right)+X(r) (\delta
    \zeta -\alpha )-2\right) + \mathcal{B}~. \label{actionL}
\end{equation}

\subsubsection{Thermodynamic Quantities from Boundary Terms}

Varying (\Ref{actionL}) with respect to $N, X, f$ the corresponding field equations read
\begin{eqnarray}
    &&2 r f'(r)+2 f(r)+\mu  X(r)^2 \left(\delta +\varepsilon  r^2\right)+X(r) (\delta  \zeta -\alpha )-2=0~,\\
    &&-\alpha +\delta  \zeta +2 \mu  X(r) \left(\delta +\varepsilon  r^2\right) =0~,\\
    &&N'(r)=0~,
\end{eqnarray}
These equations are satisfied for the solution given in equations (\ref{f}),(\ref{x}) alongside a constant $N$ which we can always set to unity with a time re-parametrization. Note, that the integral quantity in (\ref{actionL}) vanishes on-shell, and as a result $\mathcal{I}_E=\mathcal{B}$. 
In order to have a well-defined variational principle $\delta\mathcal{I}_E=0$, we have discarded a boundary term, namely
\begin{equation}
    \delta \mathcal{B} + 8\pi\beta \int_{r_h}^{\infty} dr \frac{d}{dr}\left( r \delta f \right) =0~ \to  \delta \mathcal{B} =-  8\pi \beta r \delta f(r) \big|_{r_h}^{\infty}~.
\end{equation}
It is clear from the Euclidean action that the scalar field does not introduce any boundary term. In this particular case, even if it introduced any boundary term its variation would be zero since the scalar field does not contain the black hole mass which is the only parameter allowed to vary at the boundaries. The parameters describing this black hole spacetime are the integration constant $m$ and the constants of the theory $\alpha,\delta,\zeta,\mu$. As a result, $m$ is allowed to vary and the theory-constants are fixed. For the variation of the metric function $f$, at infinity and at the event horizon we find, respectively, 
\begin{equation}
    \delta f|_\infty = -2\delta m/r~ \hspace{0.3cm}\& ~ \hspace{0.3cm} \delta f|_{r_h} = -\frac{4\pi}{\beta}\delta r_h~, \footnote{Expanding near the event horizon radius, allowing the horizon to vary and then using the definition of the inverse temperature (\ref{period}) we have: $f(r)|_{r_h}= 0 + f'(r_h)(r-r_h)+...\to \delta f(r)|_{r_h} = - f'(r_h)\delta r_h = -\frac{4\pi}{\beta}\delta r_h~$.}
\end{equation}
where we expanded near the event horizon and utilized the horizon condition $f(r_h)=0$. 
Splitting the variation of the boundary term into two parts, one at infinity and one at the event horizon 
\begin{equation}
    \delta \mathcal{B} = \delta \mathcal{B}(\infty) + \delta \mathcal{B}(r_h)
\end{equation}
we can calculate  
\begin{equation}
    \delta\mathcal{B}(\infty) = 16 \pi  \beta  \delta m ~ \hspace{0.3cm}\& ~ \hspace{0.3cm} \delta \mathcal{B}(r_h) = -32\pi^2 r_h \delta r_h = -4\pi\delta A(r_h),
\end{equation}
where $A=4\pi r_h^2$ is the area of the event horizon.

Considering the grand canonical ensemble, where the black hole is positioned in a heat bath of constant temperature $T \equiv 1/\beta$, while being capable of exchanging energy with its environment, we may integrate to obtain
\begin{equation}
    \mathcal{B}(\infty) = 16\pi\beta m~ \hspace{0.3cm}\& ~ \hspace{0.3cm}
    \mathcal{B}(r_h) = - 4\pi A(r_h)~.
\end{equation}
As a result, the action acquires the form
\begin{equation}
    \mathcal{I}_E = \mathcal{B} = 16\pi\beta m - 4\pi A(r_h) \equiv \beta \mathcal{F}~,
\end{equation}
so by comparing this with the free energy of the grand canonical ensemble $\mathcal{F} = \mathcal{M} - T \mathcal{S}$, we may identify the conserved mass and entropy of the black hole respectively as  
\begin{equation}
    \mathcal{M} = 16 \pi m~, ~ \mathcal{S} =  4\pi A(r_h)~. \label{massentropy}
\end{equation}
As a result of achieving a well-defined variational procedure, the first law of thermodynamics reads as
\begin{equation}
    \delta \mathcal{I}_E = \delta\mathcal{B}=0 \to \delta \mathcal{M} = T \delta \mathcal{S}~. \label{firstlaw}
\end{equation}

The temperature of this black hole can be obtained as
\begin{equation}
    T \equiv \frac{1}{\beta} = \frac{f'(r_h)}{4\pi} = \frac{(\alpha -\delta  \zeta )^2+8 \mu  \left(\delta
   +\varepsilon  r_h^2\right)}{32 \pi  \mu  r_h
   \left(\delta +\varepsilon  r_h^2\right)}~,
\end{equation}
where we have used the fact that $f(r_h)=0$. The coefficients in the expressions for the mass and entropy (\ref{massentropy}) are due to the coefficient of the Ricci scalar term, which as can be seen from the expression of $G_4$ has been set to unity (\ref{functions}). A coefficient of $1/16\pi $ would result in the classic definition of mass and entropy $\mathcal{M}=m~,~\mathcal{S} = A/4$. 
The existence of a non-trivial scalar field leaves the entropy (\ref{massentropy}) unaffected, since it is given by the Bekenstein-Hawking area law \cite{Bekenstein:1973ur}. The temperature, for large values of the horizon is identical to the temperature of the Schwarzschild black hole, $T_{S}=1/(4\pi r_S)$, where $r_S$ is the Schwarzchild radius, $r_S=2m$, while the temperature of small black holes is determined by the constants of the theory $T(r_h/r_S \ll 1) = ((\alpha -\delta  \zeta )^2+8 \delta  \mu)/(32
   \pi  \delta  \mu  r_h)$.
   
\begin{figure}[t]
\centering
\includegraphics[width=0.5\textwidth]{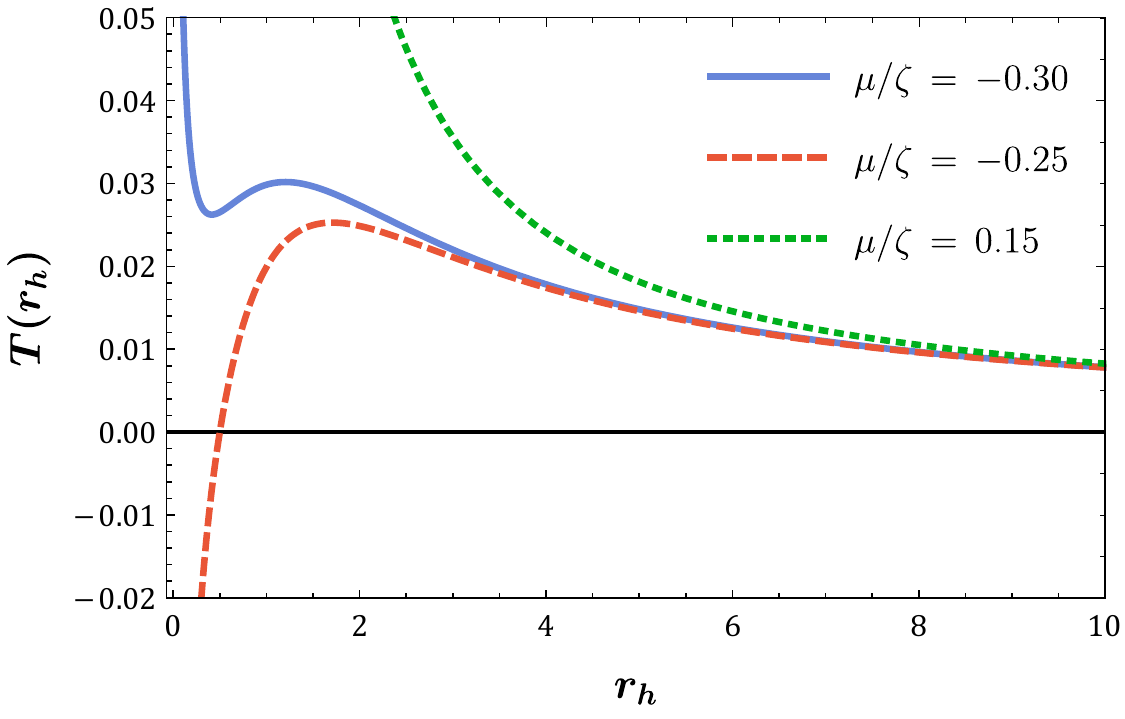}
\caption{The temperature as a function of the horizon radius. Here we have set $\delta\zeta/\alpha=4$,$\delta/\varepsilon=2$ for three different ratios of $\mu/\zeta$.}\label{temperature}
\end{figure}

Furthermore, the temperature may reach a zero value, indicating extremal black holes. In particular, the temperature vanishes for 
\begin{equation}
    r_{0} = \frac{\sqrt{-(\alpha -\delta  \zeta )^2-8 \delta  \mu
   }}{2 \sqrt{2} \sqrt{\varepsilon } \sqrt{\mu }}~.
\end{equation}
In FIG. \ref{temperature} we depict the black-hole temperature for three different scenarios. For the relations of the constants of the theory given in the caption, when $\mu/\zeta$ is positive the black hole behaves like the Schwarzschild black hole solution. Such black holes are thermally unstable when positioned in a heat bath, since they reduce in size as they get hotter. They will be described by negative heat capacity. Pedagogically, this implies that if the surrounding environment of a black hole transfers a small amount of energy to the black hole—causing its event horizon radius to expand—the black hole would cool down further. Since heat naturally flows from hotter to colder objects, the surrounding environment would continue supplying energy to the black hole. This process will lead to a self-reinforcing cycle, ultimately resulting in a runaway effect. However, when $\mu/\zeta$ is negative, the black hole may reach thermal equilibrium. As can be seen, for $\mu/\zeta=-0.25$ black holes will shrink in size as they get hotter, up to a certain size, where we encounter a phase transition and from there the black holes become colder as the horizon radius shrinks, resulting in eternal cold black holes for $r_0$. However, there is another class of solutions, those that encounter multiple phase transitions, from unstable to stable and then back to unstable, Schwarzschild-like, black hole solutions, as is evident from the case of $\mu\zeta=-0.3$. Such a behavior has been previously encountered for primary charged black hole solutions \cite{Bakopoulos:2024ogt}. This implies that the phenomenon of multiple phase transitions may be a more global one in shift and parity symmetric Horndeski theories, not particularly present in the primary-charged black hole solutions. 

\subsection{Parity Violating Theories}

In this subsection we will discuss the thermodynamics of a black hole solution arising when the parity breaking term of $F_5(X)$ is present. As an example we will examine an analytic solution that has already been obtained in \cite{Bakopoulos:2022csr}. For $F_4(X)=0=G_5(X)$ and the coupling functions 
\begin{eqnarray}
    &&G_2(X)=\eta ^2 X+\rho  \sqrt{-X}~,\\
    &&G_4(X)=\sqrt{-\alpha  X}~,\\
    &&F_5(X) = \frac{\gamma }{3 (-2 X^{5/2})}~,
\end{eqnarray}
where $\alpha,\gamma,\eta,\rho$ are coupling constants of the theory we can obtain the following solution for the metric function $f$ and the kinetic term $X$ utilizing (\ref{lorenttzianmetric})
\begin{eqnarray}
    &&f(r) = \frac{1}{4} \left(\frac{8 r^3}{l^3}+\frac{12 (r-2
   m)}{l}-\frac{3 l}{2 r}\right)^{2/3}~,\\
   &&X(r) = -\frac{\gamma  \left(l^2+4 r^2\right)^2}{4 \eta ^2
   l^3 r^4}~,
\end{eqnarray}
where $m$ is the ADM mass of the black hole and $\rho=\frac{4 \sqrt{\gamma } \eta }{l^{3/2}}{l^2},~\alpha=\frac{1}{4} \gamma  \eta ^2 l$. Here the coupling constant $\rho$ of the theory is fine-tuned in order to avoid an angle deficit at asymptotic infinity and we have redefined the constant $\alpha$ so $f(r)$ asymptotes to
\begin{equation}
    f(r\to\infty) \sim  1+\frac{r^2}{l^2}-\frac{2 m}{r}-\frac{3 l^2}{8 r^2}~,
\end{equation}
thus describing black holes in asymptotically Anti-de-Sitter (AdS) spaces. The kinetic term $X$ is finite everywhere and determined solely by the constants of the theory.

We follow the previously-described procedure, where on-shell the reduced Hamiltonian vanishes and the Euclidean action is determined by the boundary term. 
In order to have a well-defined variational principle $\delta\mathcal{I}_E=0$, we have discarded a boundary term, namely
\begin{equation}
    \delta \mathcal{B} - 8\pi\beta \gamma \int_{r_h}^{\infty} dr \frac{d}{dr}\left( \sqrt{f(r)} \delta f \right) =0~ \to  \delta \mathcal{B} =+  8\pi \beta \gamma \sqrt{f(r)} \delta f(r) \big|_{r_h}^{\infty}~.
\end{equation}
$m$ is a pure integration constant which we are allowed to vary. It is clear that the boundary term is sourced by the $F_5(X)$ part of the theory. As a result, at infinity, we find that
\begin{equation}
    \delta\mathcal{B} = \frac{16\pi\beta\gamma}{l}\delta m~.
\end{equation}
However, at the event horizon of the black hole we have $f(r_h)=0$ and hence no boundary term is generated at the event horizon of the black hole, and by comparing with the previous discussion this will imply that these solutions have a constant entropy ($\delta \mathcal{S}=0$). Consequently, a well-defined variational principle implies
\begin{equation}
\delta\mathcal{I}_E=\delta\mathcal{B}=0 \to \frac{16\pi\beta\gamma}{l}\delta m=0~,
\end{equation}
where, the constants $\beta,l$ are finite, as well as the inverse temperature $\beta$, defined by (\ref{period}) (at least so far),
and as a result, the theory predicts black hole solutions with fixed mass $m$, which is not allowed to vary: $\delta m=0$. As a result, a meaningful first law of thermodynamics (\ref{firstlaw}) would imply that $T=0$ and the black holes do not radiate. Then, according to the third law of thermodynamics, a system of zero temperature will be in its ground state having zero entropy, which is consistent with the fact that we do not obtain a boundary term at the horizon. Note, that this case is different from the Reissner-Nordstrom case, where the black holes reach thermal equilibrium in a dynamical manner, having in general a non-zero, positive temperature (and a non-zero positive entropy). 

\section{General thermodynamical remarks}\label{sec4}

As we have seen so far, the method we utilize relies on the attainment of a true extremum of the action under the fields considered in each case. This is our sole requirement, which fixes the variation of the boundary term $\delta\mathcal{B}$ that is introduced in order to achieve $\delta\mathcal{I}_E=0$. After this is dealt, we invoke the grand canonical ensemble, keeping the temperature fixed, in order to perform the integrations and the comparison with the free energy of the grand canonical ensemble will yield the mass and entropy of the black hole spacetime. In our work we deal with radially dependent scalar field, a framework that excludes black holes carrying primary scalar charge \cite{Bakopoulos:2023fmv,Baake:2023zsq,Bakopoulos:2024ogt, Bakopoulos:2023sdm}. We will actually prove that this is the case, at least for homogeneous spacetimes, in this section. This implies that there does not exist any independent integration constant that determines the behavior of the scalar field. To be as general as possible while being in concordance with the formalism developed so far, we will consider a two-degree of freedom metric, alongside the lapse function $N$ in Euclidean spacetime of the form
\begin{equation}
    ds^2 = N(r)^2 h(r) d\tau^2 + \frac{dr^2}{f(r)} + r^2 d\Omega^2~. \label{generalmetric}
\end{equation}
We will use the line element of (\ref{generalmetric}), in order to be in agreement with our analysis thus far, but also to be capable of characterizing more general cases like \cite{Rinaldi:2012vy}.
Then, following the procedure explained in detail in Section \ref{thermoexample}, we find that the boundary term related to the kinetic term of the scalar field reads
\begin{align}
    \text{boundary of $X$} =& \bigg\{-4 \pi  \beta  \sqrt{h} N \delta X \Big( \sqrt{2} s \sqrt{-X} \left(f
   \left(24 X^2 F_{5 X}+60 F_5 X-2 X G_{5 XX}-3 G_{5 X}\right)+r^2
   G_{3 X}+G_{5 X}\right)\nonumber \\
    &+4 \sqrt{f} r \left(2 X \left(2 X F_{4 X}+G_{4
   XX}\right)+G_{4 X}\right)+32 \sqrt{f} F_4 r X\Big)\bigg\}\bigg|_{r_h}^{\infty}~. \label{boundary_of_X}
\end{align}
%
%
The event horizon in our formalism is defined by $h(r_h)=0=f(r_h)$ with $\lim_{r\to r_h}\left(h(r)/f(r)\right) = \text{constant}$. In cases where $X$ is independent of the horizon radius (or equivalently the black hole mass, due to the absence of any primary hair), we have $\delta X =0$ both at infinity and at the event horizon. As a result, $X$ does not contribute to boundary terms. The independence of $X$ from the horizon radius is a common feature, as observed in Section \ref{thermoexample}. However, this does not necessarily hold in the general case. We will revisit the boundary term associated with $X$ in a later discussion.
The boundary term associated with $f$ is given by (while no boundary term arises for $h$):
\begin{align}
    \text{boundary of $f$} =& \bigg(-4 \pi  \beta    N \delta f \left(-2 G_4 r+2 X \left(
   \sqrt{2} \sqrt{f} s \sqrt{-X} \left(12 F_5 X-G_{5 X}\right)+4 F_4 r X+2
   r G_{4 X}\right)\right)\frac{\sqrt{h}}{\sqrt{f}}\bigg)\bigg|_{r_h}^{\infty}~. \label{generalentropy}
\end{align}
These boundary terms are also constrained by the field equations from which they are obtained, along with the non-ghost condition (\ref{noghost}). The expressions of the field equations are not very illuminating, and thus we skip giving their form here. However, utilizing the field equations, we may write the boundary term of $X$ as
\begin{equation}
    \text{boundary of $X$} = \frac{4 \pi  \beta  \delta X \sqrt{h} \left(f h'-f' h\right) \left(-2 
   \sqrt{2} f^{3/2} s (-X)^{3/2} \left(12 F_5 X-G_{5 X}\right)+G_2 r^3+2 G_4
   r\right)}{f^{3/2} X \left(r h'+h\right)}~.
\end{equation}
 
Interestingly, for homogeneous metrics, which satisfy $h(r)=f(r)$, the first parenthesis vanishes, since $f h'-f' h=0$ so a boundary term for $X$ does not survive. In the non-homogeneous case, i.e. $h \neq f$, the boundary term of $X$ may vanish when the metric function $f$ satisfies 
\begin{equation}
    f^{3/2} = \frac{G_2 r^3+2 G_4 r}{2 \sqrt{2} s (-X)^{3/2} \left(12 F_5 X-G_{5
   X}\right)}~. \label{constraintf}
\end{equation}
The vanishing of the boundary term related to $X$ in the case of homogeneous spacetimes implies the absence of any conserved parameter related to the scalar field $\phi$, i.e a primary scalar hair, when $h(r)=f(r)$. 
However, in the case where the scalar field is linearly time-dependent, a nonzero conjugate momentum term arises when moving to the Hamiltonian formulation \cite{Bakopoulos:2024ogt,Bakopoulos:2024zke}, resulting to a non-zero boundary term for $X$. Hence, a linearly time-dependent scalar field is the only way that one can generate a black hole carrying primary scalar hair in the context of homogeneous metric, and the corresponding field has to have a non-zero conjugate momentum \cite{Bakopoulos:2023sdm,Bakopoulos:2024zke}, just like the case of electric fields in electromagnetism \cite{Hassaine:2007py,Gonzalez:2009nn,Martinez:2005di}. Regardless of the fact of the vanishing of the boundary term of $X$, a primary scalar hair has to be characterized by a non-trivial Noether current, which cannot be the case of purely radial scalar field, as it has been shown in \cite{Bakopoulos:2023sdm}. 

In the following, we will set $N(r)=1.$, since the introduction of the lapse function $N$ was to be able to write the reduced Hamiltonian in a similar manner to the previous examples, and now there's no reason to keep two different function for the same degree of freedom ($g_{tt}$).

\subsection{Homogeneous Spacetimes}

For homogeneous spacetimes, when $h(r)=f(r)$, 
from (\ref{generalentropy}) it is possible to derive the expression for the variation of the entropy of the black hole in the context of shift-symmetric Horndeski and beyond-Horndeski theories, once the coupling functions $G_i,F_i$ are given. In particular, it will be given by
\begin{equation}
    \delta\mathcal{S} = \bigg(16 \pi^2      \left(2 G_4 r_h-2 X \left(
   \sqrt{2} \sqrt{f} s \sqrt{-X} \left(12 F_5 X-G_{5 X}\right)+4 F_4 r_h X+2
   r_h G_{4 X}\right)\right)\bigg)\delta r_h~.\label{entropyformula}
\end{equation}
To derive this result, we have used the fact that $\delta f = -4\pi \delta r_h/\beta$, utilized the grand canonical ensemble (fixed $\beta$) and required $\delta\mathcal{I}_E=0$.  All functions are evaluated at the event horizon of the black hole, and the entropy is obtained by integrating (\ref{entropyformula}) with respect to the horizon and adjusting the integration constant that emerges so that the entropy vanishes in the absence of a black hole. Note here the absence of $G_2,~ G_3$ in this expression, as well as that parity breaking terms may contribute to the total entropy, as long as $X(r_h) \neq 0$ and $\sqrt{f}  \sqrt{-X} \left(12 F_5 X-G_{5 X}\right) \neq 0$. When $F_5=G_{5X}/(12X)$ there is no contribution from the parity breaking terms and taking this into account, the no-ghost condition (\ref{noghost}) implies that 
\begin{equation}
    F_4 = + \frac{G_4}{4X^2} -\frac{G_{4X}}{2X}~,
\end{equation}
which fixes the relation between $F_4$ and $G_4$. Substituting this result back in the entropy we have the vanishing of (\ref{entropyformula}), and as a result a vanishing entropy. Hence, the relation of 
\begin{equation}
    F_5= \frac{G_{5X}}{12X}~,
\end{equation}
implies eternal black holes with a fixed mass, zero temperature and vanishing entropy according to the third law of thermodynamics. Using the fact that the variation of the metric function at infinity for any asymptotically flat black hole will be given by $\delta f = -2\delta m/r + \mathcal{O}(r^{-n})$ with $n>1$ and $m$ being the ADM mass (where the $ \mathcal{O}(r^{-n})$ terms are assumed independent of the ADM mass), the variation of the conserved black hole mass will be evaluated by utilizing (\ref{generalentropy}) as
\begin{equation}
    \delta \mathcal{M} = \lim_{r\to\infty} \Bigg(\frac{8 \pi     }{r}  \left(2 G_4 r-2 X \left(
   \sqrt{2} \sqrt{f} s \sqrt{-X} \left(12 F_5 X-G_{5 X}\right)+4 F_4 r X+2
   r G_{4 X}\right)\right)\Bigg)\delta m~. \label{massvariation}
\end{equation}
The mass may be obtained by evaluating the above limit and then integrating with respect to $m$. (In the case where the parameter $m$ is not an independent integration constant, but has somehow been promoted to a constant of the theory appearing in the action, the mass and therefore the location of event horizon are both fixed by the theory. In such a non-physical scenario, the black holes will not radiate, thus having $T=0$ and according to the third law of thermodynamics will be described by zero entropy.)
Regarding the first law of thermodynamics, such theories will satisfy the following first law
\begin{equation}
    \delta \mathcal{M} = T\delta \mathcal{S}~,
\end{equation}
where the mass and entropy entering this relation will be given by (\ref{massvariation}) and (\ref{entropyformula}) while the temperature may be found utilizing (\ref{period}).

\section{Specific Examples}\label{sec5}

Let us now proceed by discussing some particularly interesting examples, through the formalism developed thus far. 

\subsection{General Relativity}

In the case of GR we have $G_4 = 1$ while all other $G_i,F_i$ are zero and $X=0$. Hence, for the Schwarzchild black hole $f=h=1-2m/r~,N=1$ the boundary term (\ref{generalentropy}) becomes
\begin{equation}
    \text{boundary of $f$} = 8 \pi  \beta  \delta f r\Big|_{r_h}^{\infty}~.
\end{equation}
In order to cancel this term (we require $\delta \mathcal{I}_E=0$), and according to our discussion so far we have to introduce at the horizon via $\delta\mathcal{B}$ the boundary contribution
\begin{equation}
    \delta \mathcal{B}(r_h) = -4\pi \delta A \to \mathcal{B}(r_h) = -4\pi  A~.
\end{equation}
At infinity, we find
\begin{equation}
    \text{boundary of $f$} = -16\pi \beta \delta m~,
\end{equation}
and as we did before, we have to introduce 
\begin{equation}
    \delta \mathcal{B}(\infty) = 16\pi\beta \delta m \to \mathcal{B}(\infty) = 16\pi\beta M~,
\end{equation}
where the grand canonical ensemble has been evoked. As a result, using (\ref{action_free_energy}), $\mathcal{I}_E = \beta \mathcal{F} =  \mathcal{B} = \beta(16\pi m- 4\pi A/\beta)$, which in return implies $\mathcal{M} =16\pi m~, \mathcal{S} = 4\pi A$ for the mass and entropy of the black hole respectively. We would obtain the exact same result for any black hole spacetime with $G_5=F_4=F_5=0$, for example, for the asymptotically black hole carrying secondary scalar hair presented in \cite{Babichev:2017guv}.

\subsection{Shift and parity symmetric case}

As we have already discussed,  the action of parity and shift symmetric beyond Horndeski theories will be given by (\ref{beyondactionparity}). We will also discuss these theories since it has been found that they can host black-hole solutions featuring primary scalar hair. This class of theories excludes $G_3$, $G_5$ and $F_5$ and as a result, the boundary term of $f$ reads 
\begin{equation}
    \text{boundary of $f$}=  8\pi r\beta  \delta f(G_4(X) - 4 F_4(X) X^2 - 2 X G_{4X})\Big|_{r_h}^{\infty}~. \label{boundaryofprimary}
\end{equation}
Any bare Einstein-Hilbert term in the action will be included in the $G_4(X)$ coupling function, while modifications to the area law might be sourced due to the presence of the additional scalar degree of freedom. However, for a homogeneous metric ansatz with $h(r)=f(r)$, which includes the black holes featuring primary scalar hair \cite{Bakopoulos:2023fmv,Bakopoulos:2023sdm,Baake:2023zsq,Bakopoulos:2024ogt}, the field equations themselves imply a relation between the coupling functions of the theory, namely
\begin{equation}
    F_4 = \frac{-f_4}{4X^2} + \frac{G_4}{4X^2} -\frac{G_{4X}}{2X}~, \label{F4}
\end{equation}
with $f_4$ being a constant of integration. Then, reviewing (\ref{boundaryofprimary}) under new light, we deduce that the boundary term will now be given by
\begin{equation}
    \text{boundary of $f$} = 8\pi f_4 r\beta  \delta f \Big|_{r_h}^{\infty}~, \label{boundaryentropy}
\end{equation}
which implies that the constant $f_4$ is the coefficient of the bare Eintein-Hilbert term in the action. This can be seen as follows. Having (\ref{F4}) one can easily integrate this to obtain the general form of the $G_4$ coupling function. By doing so, one obtains 
\begin{equation}
    G_4(X) = c_1 \sqrt{X} + f_4 -2\sqrt{X} \int dX F_4(X)\sqrt{X}~,
\end{equation}
and comparing with the action (\ref{beyondactionparity}) one can see that indeed $f_4$ is the coupling constant of Einstein's gravity. Concluding, a homogeneous metric implies a relation between the coupling functions of the theory, which further results in the preservation of the area law for the entropy. However, a more general metric ansatz with $h \neq f$ may not satisfy the area law, since now (\ref{F4}) will not hold. 

Interestingly, writing the field equations emanating from (\ref{beyondactionparity}) in the Einstein frame, where $G_{\mu\nu} = T_{\mu\nu}$, a one degree-of-freedom (homogeneous) metric implies that $\rho(X)=- p(X)$ \cite{Jacobson:2007tj}, where $\rho$ is the (effective) energy density and $p$ the (effective) radial pressure of the scalar field, so that the Null Energy Condition (NEC) \cite{Dorlis:2023qug}
\begin{equation}
    \rho(X) + p(X) \ge 0
\end{equation}
is saturated on the bound. As a result, any black hole solution arising in the context of shift and parity symmetric Horndeski theory that is described by $g_{tt}=1/g_{rr}$ will satisfy Bekenstein's Area Law for the entropy \cite{Bekenstein:1973ur} and preserve the NEC. This implies that there must be a deeper connection between the energy conditions and the entropy of a black hole\footnote{This observation would not have been made possible, without the many discussions we have had with Panagiotis Dorlis and Sotirios-Neilos Vlachos.}. It has also been proven \cite{Neupane:2002bf} that the positivity of the energy density is required for the positivity of the entropy for non-extremal black holes in string-inspired models of gravity.
Finally, the mass will be solely determined by the ADM mass, since the boundary term is identical to the one of the GR case. We also refer to Section \ref{thermoexample} for a particular example. In the context of Horndeski cases, where the shift and parity-preserving term $F_4$ is vanishing, $F_4=0$, like \cite{Bergliaffa:2021diw, Walia:2021emv,Yang:2024cjf,Jha:2022tdl,Gohain:2025jbz}, the choice of $G_4(X) = f_4 + c_1 \sqrt{X}$ will also preserve Bekenstein's Area Law \cite{Bekenstein:1973ur} for homogeneous spacetime metrics. 

\subsection{$4-$d Einstein-Scalar-Gauss-Bonnet-Gravity}

Let us now elaborate with the four-dimensional Einstein-Scalar-Gauss-Bonnet gravity, which corresponds to the Horndeski coupling functions \cite{Bakopoulos:2022csr} 
\begin{equation}
    G_2(X) = 8\alpha X^2~,\hspace{0.2cm} G_3(X) = -8\alpha X~, \hspace{0.3cm} G_4(X) =1+4\alpha X~, \hspace{0.3cm} G_5(X) = -4 \alpha \ln X~,
\end{equation}
while the beyond Horndeski functions are both zero $F_4=0=F_5$ and the solution is a homogeneous black hole $h(r)=f(r)$ \cite{Bakopoulos:2022csr}. In particular, the solution reads ($s=-1$),
\begin{equation}
    f(r) = 1+\frac{r^2}{2 \alpha } \left(1-\sqrt{\frac{8 \alpha  m}{r^3}+1}\right)~, \hspace{0.5cm} X(r) = -\frac{\left(\sqrt{f(r)}-1\right)^2}{2 r^2}~.   
\end{equation}
Evaluating the boundary term of $f$, we find %
\begin{equation}
    \text{boundary of $f$} = 8 \pi  \beta  \delta f \left(r + 4  \sqrt{2} \alpha  \sqrt{f} 
   \sqrt{-X}-4 \alpha  r X \right)\Big|_{r_h}^{\infty}~.
\end{equation}
At the event horizon, we have
\begin{equation}
    \mathcal{B}(r_h)= -4 \pi  A - 32 \pi ^2 \alpha  \ln \left(\frac{A}{4 \pi \ell^2}\right) =- \mathcal{S}~,
\end{equation}
where $\ell$ is an arbitrary (not determined from first principles) reference length scale and our result agrees with \cite{Lu:2020iav}. $\ell$ may be fixed so that extremal black holes will be described by zero entropy according to the third law (it has also been stated that extremal black holes have zero entropy \cite{Teitelboim:1994az, Hawking:1994ii}. There is also the famous string theory result of non-zero entropy \cite{Strominger:1996sh}, however, in general, one should regard non-extreme and extreme black holes as qualitatively different objects). We note that only $G_4$ contributes to the entropy and leads to logarithmic corrections as we will see in a following subsection, where we will discuss non-homogeneous spacetimes. Moving on to the contribution at infinity, we find the usual result of $\mathcal{B}(\infty) = 16\pi m \beta$ which will give $\mathcal{M} = 16\pi m$  for the conserved black hole mass. Finally, for interesting black-string solutions of this model, we refer to \cite{Guajardo:2023uix}.

\subsection{Parity-Breaking Theories Related to Lovelock Theory}

As our next example, we consider a theory characterized by the following coupling functions:\footnote{For $n=1$, the function $G_5$ takes the form $G_5=\frac{2 \alpha}{\gamma}\ln(X)$, while for $n\neq1$, it is given by $G_5=\frac{\alpha n (-2X)^n}{\gamma (n-1) X}$.}
\begin{align*} G_2 ={}& 2\gamma^3 c \alpha n (2n-1)\frac{\left(-2X\right)^{n+1}}{n+1},\quad G_3 ={} -2\gamma^2 c \alpha (2n-1)\left(-2X\right)^n, \\[1mm] G_4 ={}& 1+\alpha(-2X)^n,\quad G_{5X}=-\frac{4\alpha n}{\gamma}\left(-2X\right)^{n-2}, \\
F_4 ={}& \frac{\gamma c+1}{4X^2}\left(1+\alpha(1-2n)\left(-2X\right)^n\right),\quad F_5 = \frac{\gamma c+1}{3\gamma}\left(-2\alpha n\right)\left(-2X\right)^{n-3}. \end{align*}
This theory is particularly interesting due to its strong connections with higher-dimensional Lovelock gravity. In the special case $n=1$ with $\gamma c=-1$, where $G_5 \sim \ln(X)$, the theory reduces to the one studied in \cite{Lu:2020iav}, which emerges from the Kaluza-Klein reduction of the Gauss-Bonnet term in Lovelock gravity (see also \cite{Fernandes:2021dsb} for an alternative derivation). Notably, this theory is equivalent to the previous example under the redefinition $\alpha \rightarrow -2\alpha$. More generally, when $\gamma c=-1$, the beyond Horndeski terms vanish, i.e., $F_4=0=F_5$, and the theory remains within the Horndeski class. In this scenario, the parameter $n+1$ can be directly associated with the order of the Lovelock term involved in the higher-dimensional reduction. For instance, the $n=2$ case, where $G_4=1+\alpha X^2$ and $G_5=X$,  corresponds to the third-order Lovelock invariant $L_3$ which interestingly via Kaluza-Klein reduction to four dimensions gives the self-tuning Paul term \cite{Kobayashi:2019hrl} found in Fab 4 \cite{Charmousis:2011bf, Charmousis:2011ea}. 
To recover General Relativity, one must take the limit $a \to 0$, which eliminates the scalar field contributions and restores the standard Einstein-Hilbert action.

The case where $\gamma c\neq-1$ is an extension of the above theory to the beyond Horndeski realm. However, black hole solutions in this case suffer from deficits related to the $\gamma c$ term. These deficits can lead to inconsistencies in the structure of the solutions, affecting their physical viability and stability. As we will demonstrate in this section, ensuring that $\gamma c=-1$—which keeps the theory within Horndeski—is also essential for maintaining a well-defined thermodynamic mass.

The solution for the aforementioned action functionals is a homogeneous black hole $h(r)=f(r)$, where the kinetic term reads
\begin{equation}
    X = -\frac{\left(1-\sqrt{c \gamma  (1-2 n) f(r)}\right)^2}{2 c \gamma ^3 (1-2
   n) r^2}~,
\end{equation}
while the metric asymptotes to 
\begin{equation}
    f(r\to\infty) \sim -\frac{1}{c \gamma }+\frac{\lambda  \left(c \gamma ^3 (1-2 n)\right)^{-n}}{c \gamma  (1-2 n)
   (n+1) r}~,
\end{equation}
where $\lambda$ is an integration constant defining the ADM mass: $\lambda=2 c \gamma  M (n+1) (2 n-1) \left(c \gamma ^3 (1-2 n)\right)^n$. Then, by applying (\ref{entropyformula}) and (\ref{massvariation}) we obtain that the entropy and the conserved mass of this solution follow
\begin{eqnarray}
    &&\mathcal{S} = -A c \gamma  \left(4 \pi -A^{1-2 n}\frac{\alpha  16^n (1-2 n)^{1-n} \pi ^{2 n}
    c^{-n} \gamma ^{-3 n}}{n-1}\right)~,\label{eqs}\\
   &&\mathcal{M} = -16 \pi   c \gamma  M~,
\end{eqnarray}
while for $\gamma c=-1$ they reduce to
\begin{eqnarray}
    &&\mathcal{S} = A \left(4 \pi+A^{1-2
   n}\frac{\alpha  (-1)^{1-3 n} 16^n (1-2 n)^{1-n} \pi ^{2 n}  c^{2 n}}{n-1} \right)~,\\
   &&\mathcal{M} = 16 \pi  M.
\end{eqnarray}
Although an analytic expression for the metric is not available in this case, approximate solutions near the horizon and at infinity suffice for the thermodynamic analysis. Since the thermodynamic quantities depend only on contributions from these regions, the full metric solution is not required. The form of the entropy is highly dependent on the order of the Lovelock term, while the value of $\gamma c$ clearly influences both the entropy and the conserved mass of the black hole. Note that $X$ remains finite at the event horizon and from (\ref{entropyformula}) we can clearly deduce that only $G_4$ and $F_4$ contribute to the entropy formula, since the $G_5,~F_5$ contributions are multiplied by $\sqrt{f}$, which is null at the horizon. For another case of black hole thermodynamics in Lovelock theories we refer to \cite{Correa:2013bza}. From the expression of the entropy, it is clear that Eq. (\ref{eqs}) holds for $n \neq 1$. For the case $n=1$, one must first substitute $n=1$ directly into the action before performing the entropy calculation, following the same procedure as in the previous case. Moreover, $n$ is not restricted to integer values and can take any real number value. However, only integer values of $n$ have a direct connection to Lovelock theory, as they correspond to specific higher-order curvature invariants arising from dimensional reduction. For $n>1$ (for integer values), the exponent in the correction term of the entropy, $A^{1-2n}$, becomes negative. Consequently, for large black holes with $A \gg 1$, the correction term diminishes, and as the Lovelock order increases, the entropy asymptotically approaches the GR limit, $\mathcal{S} \sim A/4$. On the other hand, for small black holes with $A \ll 1$, the correction term becomes dominant over the linear term, significantly altering the entropy behavior in this regime.

\subsection{Non-homogeneous Spacetimes}

Here we will discuss the non-homogeneous spacetimes. In the non-homogeneous cases, one has to solve the theory, determine the behavior of $X$ and $f$ and then proceed with a calculation of the boundary terms. $X$ may contain the metric function $f$ and therefore the horizon radius (the black hole mass), and as a result, it is allowed to vary at the boundaries.   Since we cannot have a general result as in the homogeneous case, we will discuss a very well-known black hole solution, that of \cite{Rinaldi:2012vy, Minamitsuji:2013ura}, which are solutions of the ``kinetic coupling" model, where the Einstein tensor is coupled to the derivatives of the scalar field. For various interesting aspects of this theory we refer to \cite{Arratia:2020hoy, Caceres:2023gfa}. The thermodynamics of these solutions have been discussed in \cite{Rinaldi:2012vy, Feng:2015oea, Anabalon:2013oea,Hajian:2020dcq} following various methods. It has been observed that these methods disagree at some level. This might be a computational issue, or something more profound might be at play. The Jacobson, Kang, Myers ambiguity \cite{Jacobson:1993vj} shows that such boundary-term ambiguities in the Wald formula do not affect the entropy of stationary black holes with a regular Killing horizon. However, in scalar–tensor theories these regularity conditions may fail, e.g. due to divergences of scalar derivatives, and the Wald formula may then become inapplicable. In fact, in \cite{Hajian:2020dcq} it was argued that in this case the temperature has to be altered in order for the first law of thermodynamics to hold, a scenario based on the fact that in modified gravity theories gravitons move on an effective metric.
We will show that our general formalism leads to valid results for the mass and entropy of the black hole, while the temperature remains unaltered, as required by the Euclidean continuation and finite temperature systems. This class of theories corresponds to Horndeski theory with $G_2(X)  = X$ and $G_4(X) = m_p^2/2 + zX/(2 m_p^2)$, closely following \cite{Rinaldi:2012vy}. The black hole solution corresponds to a non-homogeneous spacetime described by 
\begin{eqnarray}
    &&h(r) = \frac{3}{4}-\frac{2 M}{m_p^2 r}+\frac{m_p^2 r^2}{12 z}-\frac{\pi 
   \sqrt{z}}{8 m_p r}+\frac{\sqrt{z} \tan
   ^{-1}\left(\frac{m_p r}{\sqrt{z}}\right)}{4 m_p
   r}~,\\
   &&f(r) = \frac{h(r) \left(m_p^2 r^2+z\right)}{z \left(r h'(r)+h(r)\right)}~,
\end{eqnarray}
where $m$ is the ADM mass, while $X$ is dynamical and given by
\begin{equation}
    X(r) = \left(\frac{2 z^2}{m_p^6 r^2}+\frac{2 z}{m_p^4}\right)^{-1}~.
\end{equation}
As one can see, $X$ is independent of the black hole parameter $m$, which implies that $\delta X=0$ at the boundaries. As a result, the only contributions at the boundary will be given by (\ref{generalentropy}). For this theory, the boundary term becomes
\begin{equation}
    \text{boundary of $f$} = \left(\frac{4 \pi  \beta  \delta  f r \sqrt{h(r)} m_p^2}{\sqrt{f(r)}}-\frac{4
   \pi  \beta  \delta  f r z \sqrt{h(r)} X(r)}{\sqrt{f(r)} m_p^2}\right)\Bigg|_{r_h}^{\infty}.
\end{equation}
Now, following closely the formalism developed through the example in section \ref{thermoexample} we can deduce that the entropy and the conserved mass of the black hole will be given by
\begin{eqnarray}
    &&\mathcal{S} =A\pi   m_p^2+4 \pi ^2 z \ln \left(\frac{A m_p^2 + 4\pi z}{4 \pi \zeta^2}\right)~,\\
   &&\mathcal{M} = 8\pi M~,
\end{eqnarray}
where $\zeta$ is an arbitrary (not determined from first principles) reference length scale.
As we can see the entropy receives logarithmic corrections due to the non-minimal coupling between the scalar field and curvature. As we saw previously in the subsection regarding the $4-d$ Einstein-Scalar-Gauss-Bonnet theory, the form of $G_4 \sim 1 + \alpha X$ yields logarithmic corrections in the framework of Horndeski theories. It is important to stress that the general result of (\ref{entropyformula}) may hold in the case of non-homogeneous spacetimes when $\delta X=0$, however, one needs to compute the boundary term of $X$ and then decide whether it vanishes or not, since there is no way to realize a priori (before solving the field equations) whether the boundary of $X$ will survive, thus contributing to the entropy.

\subsection{Three-dimensional Horndeski Theory}

In this subsection, we will briefly discuss the case of shift symmetric Horndeski theories in three spacetime dimensions. We will follow the same procedure as we did for the four-dimensional case, utilizing the ADM Euclidean metric $g_{\mu\nu} = \text{diag}(N(r)^2h(r),f(r)^{-1},r^2)$, the grand canonical ensemble, fixed $\beta$, which will be given by (\ref{period}) and we obtain the following boundary terms for $X$ and $f$
\begin{eqnarray}
    && \text{boundary of $X$} = -2 \pi  \beta  \sqrt{h} N \delta X \left(4 \sqrt{f} X G_{4 XX}+2
   \sqrt{f} G_{4 X}+ \sqrt{2} r s \sqrt{-X} G_{3 X}\right)\bigg|_{r_h}^{\infty}~,\\ 
   && \text{boundary of $f$} = 2 \pi  \beta   N\delta f  \left(G_4-2 X G_{4
   X}\right)\sqrt{h/f}\bigg|_{r_h}^{\infty}~. \label{boundaryf}
\end{eqnarray}
In  this formalism, the $\mathcal{L}_5$ term is a total derivative in three spacetime dimensions and does not appear in the reduced action, which after integration by parts takes the simple form 
\begin{equation}
    \mathcal{I}_E^{3\text{d}} = 2 \pi  \beta\int dr   \frac{\sqrt{h}}{\sqrt{f}}N \left(f' \left(G_4-2 X G_{4 X}\right)-
   \sqrt{2} \sqrt{f} r s \sqrt{-X} G_{3 X} X'-2 f X' \left(2 X G_{4
   \text{XX}}+G_{4 X}\right)-G_2 r\right) + \mathcal{B}^{3\text{d}}~.
\end{equation}
By using the field equations we may write the boundary term of $X$ as
\begin{equation}
    \text{boundary of $X$} = \frac{2 \pi  \beta  \delta X \left(f h'-h f'\right) \left(G_4-2 G_{4 X}
   X\right)}{\sqrt{f} \sqrt{h} X}~,
\end{equation}
from which is clear that for homogeneous metrics with $h=f$ or for $G_4 \sim \sqrt{X}$ the boundary term vanishes and no contribution from the scalar field survives. However, in the second case, we have also the vanishing of the boundary term of $f$ and hence no entropy or mass may be attached to this black hole. From the boundary term of $f$ (\ref{boundaryf}) we may read the entropy and mass variations of any three-dimensional homogeneous black hole spacetime as
\begin{eqnarray}
    &&\delta\mathcal{S} = 8 \pi^2      \left(G_4-2 X G_{4
   X}\right)\delta r_h~, \label{entropyvariation3d} \\
   &&\delta\mathcal{M} = \lim_{r\to\infty}\left(2 \pi     \left(G_4-2 X G_{4
   X}\right)\right)\delta m~. \label{massvariation3d}
\end{eqnarray}

Not many black hole solutions are known in the shift symmetric Horndeski theory in three dimensions. Some interesting solutions may be found in \cite{Bravo-Gaete:2014haa, Alkac:2024hvu,Hennigar:2020fkv}. There are many examples in the non shift-symmetric Horndeski theory, including black holes with conformally (\cite{Martinez:1996gn, Cardenas:2014kaa}) or minimally coupled scalar fields with a self-interacting potential \cite{Priyadarshinee:2023cmi, Sardeshpande:2024bnk, Karakasis:2021lnq, Karakasis:2023ljt}. A well-known solution in the framework of shift-symmetric Horndeski theories is the one presented in \cite{Bravo-Gaete:2014haa}, which corresponds to a Horndeski theory with $G_4=1 + \eta X/2$ and $G_2 =-2\Lambda + \alpha X$. The solution to the resulting field equations is a homogeneous black hole with $f(r) = r^2/\ell^2 -m$ and $X=(\Lambda  l^2+1)/\eta$ with $\ell^2 = \eta/\alpha$ and $m$ an integration constant. Then, using (\ref{entropyvariation3d}) and (\ref{massvariation3d}) we may readily integrate and obtain the entropy and mass respectively as 
\begin{eqnarray}
    &&\mathcal{S} = 8\pi^2\left(\frac{1-\Lambda  l^2}{2} \right)r_h~,\\
    &&\mathcal{M} = 2\pi m \left(\frac{1-\Lambda  l^2}{2} \right)~,
\end{eqnarray}
and our result agrees with the one presented in \cite{Bravo-Gaete:2014haa}.

\section{Conclusions}\label{conc}

Scalar-tensor theories, and in particular Horndeski gravity, provide one of the most general and well-motivated extensions of General Relativity that remain free from ghost-like instabilities. These theories introduce non-minimal couplings between a scalar field and curvature, leading to a rich spectrum of gravitational phenomena, including novel black hole solutions, modifications to cosmological evolution, and deviations from GR predictions in the strong-field regime. In recent years, Horndeski and beyond-Horndeski theories have played a crucial role in exploring alternative gravity models, offering a framework to address fundamental questions in gravitational physics, such as the nature of dark energy, the structure of compact objects, and the stability of black hole solutions. 

A particularly important subclass of these theories is shift-symmetric Horndeski gravity, where the scalar field enters the action only through its derivatives. This additional symmetry not only simplifies the field equations but also plays a crucial role in the derivation of exact solutions. Many known black hole solutions in Horndeski gravity, including those with primary scalar hair, rely on the presence of shift symmetry to evade standard no-hair theorems. Moreover, certain shift-symmetric theories naturally arise from the compactification of higher-dimensional Lovelock gravity, reinforcing their theoretical motivation. Given the strong connections between shift symmetry, exact black hole solutions, and higher-dimensional gravity, understanding the thermodynamic properties of these models is a necessary step in assessing their physical relevance.

Black hole thermodynamics provides a powerful tool for probing gravitational theories beyond General Relativity, revealing deep connections between classical gravity, quantum effects, and statistical mechanics. The formulation of black hole mechanics as a thermodynamic system has led to profound insights into the nature of entropy, the information problem, and the role of quantum corrections in gravitational physics. In modified gravity, black hole thermodynamics is particularly important for understanding how additional degrees of freedom influence the fundamental laws of gravitational interactions. In Horndeski theories, where the scalar field dynamically couples to curvature, thermodynamic properties such as entropy, temperature, and mass can differ significantly from their GR counterparts.

While the Wald entropy formula has been widely used to derive black hole entropy in modified gravity theories, it is known to suffer from significant ambiguities in the context of scalar-tensor models such as Horndeski and beyond-Horndeski theories. These ambiguities stem from the presence of covariant derivatives of the scalar field in the action and the dependence of the entropy on the specific form of the Lagrangian — especially on whether integration by parts is performed before or after taking functional derivatives with respect to the Riemann tensor. Since equivalent actions (leading to the same equations of motion) can differ by boundary terms, the Wald procedure may yield inequivalent results, and even differ between implementations. As shown in several works, these discrepancies can lead to violations of the first law of thermodynamics unless ad hoc modifications are introduced.

In this work, we have investigated the thermodynamics of black hole solutions in shift-symmetric Horndeski and beyond-Horndeski theories, focusing on the role of shift and parity symmetries in the thermodynamic properties of black holes. By employing the Euclidean method, we developed a systematic framework for extracting black hole entropy, temperature, and mass directly from the action, ensuring a well-defined variational principle. Our approach provides a unified methodology for analyzing black hole thermodynamics in a broad class of modified gravity theories. A key result of our analysis is the derivation of a general expression for the variation of entropy in shift-symmetric beyond-Horndeski theories. This expression allows for the computation of black hole entropy by integrating with respect to the event horizon radius, ensuring that it vanishes in the absence of a black hole.

We examined specific black hole solutions in shift- and parity-symmetric theories, as well as cases where parity symmetry is explicitly broken, revealing distinct thermodynamic behaviors in these two regimes. In the shift- and parity-symmetric case, we found that the entropy follows the standard Bekenstein-Hawking area law, consistent with previous results in the literature, regardless of the explicit form of the coupling functions. This suggests that the presence of shift and parity symmetry preserves key thermodynamic properties of black holes, even in modified gravity settings. In contrast, when parity symmetry is broken, we identified cases where the entropy vanishes, indicating a fundamental departure from standard thermodynamic expectations. Such solutions exhibit fixed mass and zero temperature, making them thermodynamically non-radiating, a phenomenon that could have implications for the stability and long-term evolution of these black holes.

Our results also provide new perspectives on the broader implications of black hole thermodynamics in modified gravity. The connection between higher-dimensional Lovelock gravity and certain Horndeski subclasses further strengthens the motivation for studying these models, as they offer a low-energy effective description of fundamental gravitational theories.
Additionally, our analysis clarifies the limitations of previous entropy calculations within Horndeski gravity, where the presence of covariant derivatives of the scalar field in the action can lead to inconsistencies when applying the Wald entropy formula. These inconsistencies become particularly manifest in non-homogeneous spacetimes, where the Wald formula fails to yield the correct entropy and may even violate the first law of thermodynamics. As emphasized in \cite{Jacobson:1993vj}, such ambiguities do not affect stationary black holes with a regular Killing horizon, but in scalar–tensor theories these regularity assumptions may fail, e.g. due to divergences of scalar derivatives at the horizon \cite{Hajian:2020dcq}, and the Wald formula can then become ill-defined. In contrast, the Euclidean method we employ circumvents these issues by systematically accounting for the boundary contributions necessary for a well-defined variational principle. By explicitly constructing the appropriate boundary terms, our approach ensures thermodynamic consistency and extends the applicability of black hole thermodynamics in scalar-tensor and beyond-Horndeski theories.

\section{Acknowledgements}
\noindent A.B. acknowledges participation in the COST Association Action CA21136 “Addressing observational tensions in cosmology with systematics and fundamental physics (CosmoVerse)”. We acknowledge useful discussions with Panos Dorlis, Sotirios-Neilos Vlachos, Theodoros Nakas and Nikos Chatzifotis.

\bibliography{Refs}

\providecommand{\href}[2]{#2}\begingroup\raggedright\begin{thebibliography}{100}

\bibitem{Horndeski:1974wa}
G.~W. Horndeski, ``{Second-order scalar-tensor field equations in a four-dimensional space},'' \href{http://dx.doi.org/10.1007/BF01807638}{{\em Int. J. Theor. Phys.} {\bfseries 10} (1974) 363--384}.

\bibitem{Nicolis:2008in}
A.~Nicolis, R.~Rattazzi, and E.~Trincherini, ``{The Galileon as a local modification of gravity},'' \href{http://dx.doi.org/10.1103/PhysRevD.79.064036}{{\em Phys. Rev. D} {\bfseries 79} (2009) 064036}, \href{http://arxiv.org/abs/0811.2197}{{\ttfamily arXiv:0811.2197 [hep-th]}}.

\bibitem{Deffayet:2009mn}
C.~Deffayet, S.~Deser, and G.~Esposito-Farese, ``{Generalized Galileons: All scalar models whose curved background extensions maintain second-order field equations and stress-tensors},'' \href{http://dx.doi.org/10.1103/PhysRevD.80.064015}{{\em Phys. Rev. D} {\bfseries 80} (2009) 064015}, \href{http://arxiv.org/abs/0906.1967}{{\ttfamily arXiv:0906.1967 [gr-qc]}}.

\bibitem{Deffayet:2009wt}
C.~Deffayet, G.~Esposito-Farese, and A.~Vikman, ``{Covariant Galileon},'' \href{http://dx.doi.org/10.1103/PhysRevD.79.084003}{{\em Phys. Rev. D} {\bfseries 79} (2009) 084003}, \href{http://arxiv.org/abs/0901.1314}{{\ttfamily arXiv:0901.1314 [hep-th]}}.

\bibitem{Deffayet:2011gz}
C.~Deffayet, X.~Gao, D.~A. Steer, and G.~Zahariade, ``{From k-essence to generalised Galileons},'' \href{http://dx.doi.org/10.1103/PhysRevD.84.064039}{{\em Phys. Rev. D} {\bfseries 84} (2011) 064039}, \href{http://arxiv.org/abs/1103.3260}{{\ttfamily arXiv:1103.3260 [hep-th]}}.

\bibitem{Kanti:1995vq}
P.~Kanti, N.~E. Mavromatos, J.~Rizos, K.~Tamvakis, and E.~Winstanley, ``{Dilatonic black holes in higher curvature string gravity},'' \href{http://dx.doi.org/10.1103/PhysRevD.54.5049}{{\em Phys. Rev. D} {\bfseries 54} (1996) 5049--5058}, \href{http://arxiv.org/abs/hep-th/9511071}{{\ttfamily arXiv:hep-th/9511071}}.

\bibitem{Antoniou:2017acq}
G.~Antoniou, A.~Bakopoulos, and P.~Kanti, ``{Evasion of No-Hair Theorems and Novel Black-Hole Solutions in Gauss-Bonnet Theories},'' \href{http://dx.doi.org/10.1103/PhysRevLett.120.131102}{{\em Phys. Rev. Lett.} {\bfseries 120} no.~13, (2018) 131102}, \href{http://arxiv.org/abs/1711.03390}{{\ttfamily arXiv:1711.03390 [hep-th]}}.

\bibitem{Doneva:2017bvd}
D.~D. Doneva and S.~S. Yazadjiev, ``{New Gauss-Bonnet Black Holes with Curvature-Induced Scalarization in Extended Scalar-Tensor Theories},'' \href{http://dx.doi.org/10.1103/PhysRevLett.120.131103}{{\em Phys. Rev. Lett.} {\bfseries 120} no.~13, (2018) 131103}, \href{http://arxiv.org/abs/1711.01187}{{\ttfamily arXiv:1711.01187 [gr-qc]}}.

\bibitem{Silva:2017uqg}
H.~O. Silva, J.~Sakstein, L.~Gualtieri, T.~P. Sotiriou, and E.~Berti, ``{Spontaneous scalarization of black holes and compact stars from a Gauss-Bonnet coupling},'' \href{http://dx.doi.org/10.1103/PhysRevLett.120.131104}{{\em Phys. Rev. Lett.} {\bfseries 120} no.~13, (2018) 131104}, \href{http://arxiv.org/abs/1711.02080}{{\ttfamily arXiv:1711.02080 [gr-qc]}}.

\bibitem{Antoniou:2017hxj}
G.~Antoniou, A.~Bakopoulos, and P.~Kanti, ``{Black-Hole Solutions with Scalar Hair in Einstein-Scalar-Gauss-Bonnet Theories},'' \href{http://dx.doi.org/10.1103/PhysRevD.97.084037}{{\em Phys. Rev. D} {\bfseries 97} no.~8, (2018) 084037}, \href{http://arxiv.org/abs/1711.07431}{{\ttfamily arXiv:1711.07431 [hep-th]}}.

\bibitem{Deffayet:2010qz}
C.~Deffayet, O.~Pujolas, I.~Sawicki, and A.~Vikman, ``{Imperfect Dark Energy from Kinetic Gravity Braiding},'' \href{http://dx.doi.org/10.1088/1475-7516/2010/10/026}{{\em JCAP} {\bfseries 10} (2010) 026}, \href{http://arxiv.org/abs/1008.0048}{{\ttfamily arXiv:1008.0048 [hep-th]}}.

\bibitem{Pujolas:2011he}
O.~Pujolas, I.~Sawicki, and A.~Vikman, ``{The Imperfect Fluid behind Kinetic Gravity Braiding},'' \href{http://dx.doi.org/10.1007/JHEP11(2011)156}{{\em JHEP} {\bfseries 11} (2011) 156}, \href{http://arxiv.org/abs/1103.5360}{{\ttfamily arXiv:1103.5360 [hep-th]}}.

\bibitem{Kobayashi:2010cm}
T.~Kobayashi, M.~Yamaguchi, and J.~Yokoyama, ``{G-inflation: Inflation driven by the Galileon field},'' \href{http://dx.doi.org/10.1103/PhysRevLett.105.231302}{{\em Phys. Rev. Lett.} {\bfseries 105} (2010) 231302}, \href{http://arxiv.org/abs/1008.0603}{{\ttfamily arXiv:1008.0603 [hep-th]}}.

\bibitem{Appleby:2011aa}
S.~Appleby and E.~V. Linder, ``{The Paths of Gravity in Galileon Cosmology},'' \href{http://dx.doi.org/10.1088/1475-7516/2012/03/043}{{\em JCAP} {\bfseries 03} (2012) 043}, \href{http://arxiv.org/abs/1112.1981}{{\ttfamily arXiv:1112.1981 [astro-ph.CO]}}.

\bibitem{Creminelli:2008wc}
P.~Creminelli, G.~D'Amico, J.~Norena, and F.~Vernizzi, ``{The Effective Theory of Quintessence: the w\ensuremath{<}-1 Side Unveiled},'' \href{http://dx.doi.org/10.1088/1475-7516/2009/02/018}{{\em JCAP} {\bfseries 02} (2009) 018}, \href{http://arxiv.org/abs/0811.0827}{{\ttfamily arXiv:0811.0827 [astro-ph]}}.

\bibitem{Babichev:2013cya}
E.~Babichev and C.~Charmousis, ``{Dressing a black hole with a time-dependent Galileon},'' \href{http://dx.doi.org/10.1007/JHEP08(2014)106}{{\em JHEP} {\bfseries 08} (2014) 106}, \href{http://arxiv.org/abs/1312.3204}{{\ttfamily arXiv:1312.3204 [gr-qc]}}.

\bibitem{Babichev:2016rlq}
E.~Babichev, C.~Charmousis, and A.~Leh\'ebel, ``{Black holes and stars in Horndeski theory},'' \href{http://dx.doi.org/10.1088/0264-9381/33/15/154002}{{\em Class. Quant. Grav.} {\bfseries 33} no.~15, (2016) 154002}, \href{http://arxiv.org/abs/1604.06402}{{\ttfamily arXiv:1604.06402 [gr-qc]}}.

\bibitem{Minamitsuji:2016hkk}
M.~Minamitsuji and H.~O. Silva, ``{Relativistic stars in scalar-tensor theories with disformal coupling},'' \href{http://dx.doi.org/10.1103/PhysRevD.93.124041}{{\em Phys. Rev. D} {\bfseries 93} no.~12, (2016) 124041}, \href{http://arxiv.org/abs/1604.07742}{{\ttfamily arXiv:1604.07742 [gr-qc]}}.

\bibitem{Charmousis:2012dw}
C.~Charmousis, B.~Gouteraux, and E.~Kiritsis, ``{Higher-derivative scalar-vector-tensor theories: black holes, Galileons, singularity cloaking and holography},'' \href{http://dx.doi.org/10.1007/JHEP09(2012)011}{{\em JHEP} {\bfseries 09} (2012) 011}, \href{http://arxiv.org/abs/1206.1499}{{\ttfamily arXiv:1206.1499 [hep-th]}}.

\bibitem{Lecoeur:2024kwe}
N.~Lecoeur, {\em {Exact black hole solutions in scalar-tensor theories}}.
\newblock PhD thesis, U. Paris-Saclay, 2024.
\newblock \href{http://arxiv.org/abs/2406.11095}{{\ttfamily arXiv:2406.11095 [gr-qc]}}.

\bibitem{Lu:2020iav}
H.~Lu and Y.~Pang, ``{Horndeski gravity as $D \rightarrow 4$ limit of Gauss-Bonnet},'' \href{http://dx.doi.org/10.1016/j.physletb.2020.135717}{{\em Phys. Lett. B} {\bfseries 809} (2020) 135717}, \href{http://arxiv.org/abs/2003.11552}{{\ttfamily arXiv:2003.11552 [gr-qc]}}.

\bibitem{Glavan:2019inb}
D.~Glavan and C.~Lin, ``{Einstein-Gauss-Bonnet Gravity in Four-Dimensional Spacetime},'' \href{http://dx.doi.org/10.1103/PhysRevLett.124.081301}{{\em Phys. Rev. Lett.} {\bfseries 124} no.~8, (2020) 081301}, \href{http://arxiv.org/abs/1905.03601}{{\ttfamily arXiv:1905.03601 [gr-qc]}}.

\bibitem{Fernandes:2021dsb}
P.~G.~S. Fernandes, ``{Gravity with a generalized conformal scalar field: theory and solutions},'' \href{http://dx.doi.org/10.1103/PhysRevD.103.104065}{{\em Phys. Rev. D} {\bfseries 103} no.~10, (2021) 104065}, \href{http://arxiv.org/abs/2105.04687}{{\ttfamily arXiv:2105.04687 [gr-qc]}}.

\bibitem{Alkac:2022fuc}
G.~Alkac, G.~D. Ozen, and G.~Suer, ``{Lower-dimensional limits of cubic Lovelock gravity},'' \href{http://dx.doi.org/10.1016/j.nuclphysb.2022.116027}{{\em Nucl. Phys. B} {\bfseries 985} (2022) 116027}, \href{http://arxiv.org/abs/2203.01811}{{\ttfamily arXiv:2203.01811 [gr-qc]}}.

\bibitem{Sotiriou:2013qea}
T.~P. Sotiriou and S.-Y. Zhou, ``{Black hole hair in generalized scalar-tensor gravity},'' \href{http://dx.doi.org/10.1103/PhysRevLett.112.251102}{{\em Phys. Rev. Lett.} {\bfseries 112} (2014) 251102}, \href{http://arxiv.org/abs/1312.3622}{{\ttfamily arXiv:1312.3622 [gr-qc]}}.

\bibitem{Sotiriou:2014pfa}
T.~P. Sotiriou and S.-Y. Zhou, ``{Black hole hair in generalized scalar-tensor gravity: An explicit example},'' \href{http://dx.doi.org/10.1103/PhysRevD.90.124063}{{\em Phys. Rev. D} {\bfseries 90} (2014) 124063}, \href{http://arxiv.org/abs/1408.1698}{{\ttfamily arXiv:1408.1698 [gr-qc]}}.

\bibitem{Bakopoulos:2022csr}
A.~Bakopoulos, C.~Charmousis, P.~Kanti, and N.~Lecoeur, ``{Compact objects of spherical symmetry in beyond Horndeski theories},'' \href{http://dx.doi.org/10.1007/JHEP08(2022)055}{{\em JHEP} {\bfseries 08} (2022) 055}, \href{http://arxiv.org/abs/2203.14595}{{\ttfamily arXiv:2203.14595 [gr-qc]}}.

\bibitem{Minamitsuji:2019shy}
M.~Minamitsuji and J.~Edholm, ``{Black hole solutions in shift-symmetric degenerate higher-order scalar-tensor theories},'' \href{http://dx.doi.org/10.1103/PhysRevD.100.044053}{{\em Phys. Rev. D} {\bfseries 100} no.~4, (2019) 044053}, \href{http://arxiv.org/abs/1907.02072}{{\ttfamily arXiv:1907.02072 [gr-qc]}}.

\bibitem{Bakopoulos:2023fmv}
A.~Bakopoulos, C.~Charmousis, P.~Kanti, N.~Lecoeur, and T.~Nakas, ``{Black holes with primary scalar hair},'' \href{http://dx.doi.org/10.1103/PhysRevD.109.024032}{{\em Phys. Rev. D} {\bfseries 109} no.~2, (2024) 024032}, \href{http://arxiv.org/abs/2310.11919}{{\ttfamily arXiv:2310.11919 [gr-qc]}}.

\bibitem{Baake:2023zsq}
O.~Baake, A.~Cisterna, M.~Hassaine, and U.~Hernandez-Vera, ``{Endowing black holes with beyond-Horndeski primary hair: An exact solution framework for scalarizing in every dimension},'' \href{http://dx.doi.org/10.1103/PhysRevD.109.064024}{{\em Phys. Rev. D} {\bfseries 109} no.~6, (2024) 064024}, \href{http://arxiv.org/abs/2312.05207}{{\ttfamily arXiv:2312.05207 [hep-th]}}.

\bibitem{Bakopoulos:2023sdm}
A.~Bakopoulos, N.~Chatzifotis, and T.~Nakas, ``{Compact objects with primary hair in shift and parity symmetric beyond Horndeski gravities},'' \href{http://dx.doi.org/10.1103/PhysRevD.110.024044}{{\em Phys. Rev. D} {\bfseries 110} no.~2, (2024) 024044}, \href{http://arxiv.org/abs/2312.17198}{{\ttfamily arXiv:2312.17198 [gr-qc]}}.

\bibitem{Petronikolou:2021shp}
M.~Petronikolou, S.~Basilakos, and E.~N. Saridakis, ``{Alleviating H0 tension in Horndeski gravity},'' \href{http://dx.doi.org/10.1103/PhysRevD.106.124051}{{\em Phys. Rev. D} {\bfseries 106} no.~12, (2022) 124051}, \href{http://arxiv.org/abs/2110.01338}{{\ttfamily arXiv:2110.01338 [gr-qc]}}.

\bibitem{Gleyzes:2014dya}
J.~Gleyzes, D.~Langlois, F.~Piazza, and F.~Vernizzi, ``{Healthy theories beyond Horndeski},'' \href{http://dx.doi.org/10.1103/PhysRevLett.114.211101}{{\em Phys. Rev. Lett.} {\bfseries 114} no.~21, (2015) 211101}, \href{http://arxiv.org/abs/1404.6495}{{\ttfamily arXiv:1404.6495 [hep-th]}}.

\bibitem{BenAchour:2016cay}
J.~Ben~Achour, D.~Langlois, and K.~Noui, ``{Degenerate higher order scalar-tensor theories beyond Horndeski and disformal transformations},'' \href{http://dx.doi.org/10.1103/PhysRevD.93.124005}{{\em Phys. Rev. D} {\bfseries 93} no.~12, (2016) 124005}, \href{http://arxiv.org/abs/1602.08398}{{\ttfamily arXiv:1602.08398 [gr-qc]}}.

\bibitem{Kobayashi:2019hrl}
T.~Kobayashi, ``{Horndeski theory and beyond: a review},'' \href{http://dx.doi.org/10.1088/1361-6633/ab2429}{{\em Rept. Prog. Phys.} {\bfseries 82} no.~8, (2019) 086901}, \href{http://arxiv.org/abs/1901.07183}{{\ttfamily arXiv:1901.07183 [gr-qc]}}.

\bibitem{Gleyzes:2015pma}
J.~Gleyzes, D.~Langlois, M.~Mancarella, and F.~Vernizzi, ``{Effective Theory of Interacting Dark Energy},'' \href{http://dx.doi.org/10.1088/1475-7516/2015/08/054}{{\em JCAP} {\bfseries 08} (2015) 054}, \href{http://arxiv.org/abs/1504.05481}{{\ttfamily arXiv:1504.05481 [astro-ph.CO]}}.

\bibitem{Creminelli:2017sry}
P.~Creminelli and F.~Vernizzi, ``{Dark Energy after GW170817 and GRB170817A},'' \href{http://dx.doi.org/10.1103/PhysRevLett.119.251302}{{\em Phys. Rev. Lett.} {\bfseries 119} no.~25, (2017) 251302}, \href{http://arxiv.org/abs/1710.05877}{{\ttfamily arXiv:1710.05877 [astro-ph.CO]}}.

\bibitem{Charmousis:2019vnf}
C.~Charmousis, M.~Crisostomi, R.~Gregory, and N.~Stergioulas, ``{Rotating Black Holes in Higher Order Gravity},'' \href{http://dx.doi.org/10.1103/PhysRevD.100.084020}{{\em Phys. Rev. D} {\bfseries 100} no.~8, (2019) 084020}, \href{http://arxiv.org/abs/1903.05519}{{\ttfamily arXiv:1903.05519 [hep-th]}}.

\bibitem{Anson:2020trg}
T.~Anson, E.~Babichev, C.~Charmousis, and M.~Hassaine, ``{Disforming the Kerr metric},'' \href{http://dx.doi.org/10.1007/JHEP01(2021)018}{{\em JHEP} {\bfseries 01} (2021) 018}, \href{http://arxiv.org/abs/2006.06461}{{\ttfamily arXiv:2006.06461 [gr-qc]}}.

\bibitem{Walia:2021emv}
R.~K. Walia, S.~D. Maharaj, and S.~G. Ghosh, ``{Rotating Black Holes in Horndeski Gravity: Thermodynamic and Gravitational Lensing},'' \href{http://dx.doi.org/10.1140/epjc/s10052-022-10451-5}{{\em Eur. Phys. J. C} {\bfseries 82} (2022) 547}, \href{http://arxiv.org/abs/2109.08055}{{\ttfamily arXiv:2109.08055 [gr-qc]}}.

\bibitem{Maselli:2015tta}
A.~Maselli, P.~Pani, L.~Gualtieri, and V.~Ferrari, ``{Rotating black holes in Einstein-Dilaton-Gauss-Bonnet gravity with finite coupling},'' \href{http://dx.doi.org/10.1103/PhysRevD.92.083014}{{\em Phys. Rev. D} {\bfseries 92} no.~8, (2015) 083014}, \href{http://arxiv.org/abs/1507.00680}{{\ttfamily arXiv:1507.00680 [gr-qc]}}.

\bibitem{Kleihaus:2011tg}
B.~Kleihaus, J.~Kunz, and E.~Radu, ``{Rotating Black Holes in Dilatonic Einstein-Gauss-Bonnet Theory},'' \href{http://dx.doi.org/10.1103/PhysRevLett.106.151104}{{\em Phys. Rev. Lett.} {\bfseries 106} (2011) 151104}, \href{http://arxiv.org/abs/1101.2868}{{\ttfamily arXiv:1101.2868 [gr-qc]}}.

\bibitem{Bakopoulos:2020dfg}
A.~Bakopoulos, P.~Kanti, and N.~Pappas, ``{Large and ultracompact Gauss-Bonnet black holes with a self-interacting scalar field},'' \href{http://dx.doi.org/10.1103/PhysRevD.101.084059}{{\em Phys. Rev. D} {\bfseries 101} no.~8, (2020) 084059}, \href{http://arxiv.org/abs/2003.02473}{{\ttfamily arXiv:2003.02473 [hep-th]}}.

\bibitem{Bakopoulos:2018nui}
A.~Bakopoulos, G.~Antoniou, and P.~Kanti, ``{Novel Black-Hole Solutions in Einstein-Scalar-Gauss-Bonnet Theories with a Cosmological Constant},'' \href{http://dx.doi.org/10.1103/PhysRevD.99.064003}{{\em Phys. Rev. D} {\bfseries 99} no.~6, (2019) 064003}, \href{http://arxiv.org/abs/1812.06941}{{\ttfamily arXiv:1812.06941 [hep-th]}}.

\bibitem{Bakopoulos:2023hkh}
A.~Bakopoulos and T.~Nakas, ``{Novel exact ultracompact and ultrasparse hairy black holes emanating from regular and phantom scalar fields},'' \href{http://dx.doi.org/10.1103/PhysRevD.107.124035}{{\em Phys. Rev. D} {\bfseries 107} no.~12, (2023) 124035}, \href{http://arxiv.org/abs/2303.09116}{{\ttfamily arXiv:2303.09116 [gr-qc]}}.

\bibitem{Bakopoulos:2021dry}
A.~Bakopoulos and T.~Nakas, ``{Analytic and asymptotically flat hairy (ultra-compact) black-hole solutions and their axial perturbations},'' \href{http://dx.doi.org/10.1007/JHEP04(2022)096}{{\em JHEP} {\bfseries 04} (2022) 096}, \href{http://arxiv.org/abs/2107.05656}{{\ttfamily arXiv:2107.05656 [gr-qc]}}.

\bibitem{Karakasis:2021lnq}
T.~Karakasis, E.~Papantonopoulos, Z.-Y. Tang, and B.~Wang, ``{Black holes of (2+1)-dimensional $f(R)$ gravity coupled to a scalar field},'' \href{http://dx.doi.org/10.1103/PhysRevD.103.064063}{{\em Phys. Rev. D} {\bfseries 103} no.~6, (2021) 064063}, \href{http://arxiv.org/abs/2101.06410}{{\ttfamily arXiv:2101.06410 [gr-qc]}}.

\bibitem{Karakasis:2021rpn}
T.~Karakasis, E.~Papantonopoulos, Z.-Y. Tang, and B.~Wang, ``{Exact black hole solutions with a conformally coupled scalar field and dynamic Ricci curvature in f(R) gravity theories},'' \href{http://dx.doi.org/10.1140/epjc/s10052-021-09717-1}{{\em Eur. Phys. J. C} {\bfseries 81} no.~10, (2021) 897}, \href{http://arxiv.org/abs/2103.14141}{{\ttfamily arXiv:2103.14141 [gr-qc]}}.

\bibitem{Karakasis:2021ttn}
T.~Karakasis, E.~Papantonopoulos, Z.-Y. Tang, and B.~Wang, ``{(2+1)-dimensional black holes in f(R,\ensuremath{\phi}) gravity},'' \href{http://dx.doi.org/10.1103/PhysRevD.105.044038}{{\em Phys. Rev. D} {\bfseries 105} no.~4, (2022) 044038}, \href{http://arxiv.org/abs/2201.00035}{{\ttfamily arXiv:2201.00035 [gr-qc]}}.

\bibitem{Karakasis:2023hni}
T.~Karakasis, N.~E. Mavromatos, and E.~Papantonopoulos, ``{Regular compact objects with scalar hair},'' \href{http://dx.doi.org/10.1103/PhysRevD.108.024001}{{\em Phys. Rev. D} {\bfseries 108} no.~2, (2023) 024001}, \href{http://arxiv.org/abs/2305.00058}{{\ttfamily arXiv:2305.00058 [gr-qc]}}.

\bibitem{Karakasis:2023ljt}
T.~Karakasis, G.~Koutsoumbas, and E.~Papantonopoulos, ``{Black holes with scalar hair in three dimensions},'' \href{http://dx.doi.org/10.1103/PhysRevD.107.124047}{{\em Phys. Rev. D} {\bfseries 107} no.~12, (2023) 124047}, \href{http://arxiv.org/abs/2305.00686}{{\ttfamily arXiv:2305.00686 [gr-qc]}}.

\bibitem{Tang:2020sjs}
Z.-Y. Tang, B.~Wang, T.~Karakasis, and E.~Papantonopoulos, ``{Curvature scalarization of black holes in f(R) gravity},'' \href{http://dx.doi.org/10.1103/PhysRevD.104.064017}{{\em Phys. Rev. D} {\bfseries 104} no.~6, (2021) 064017}, \href{http://arxiv.org/abs/2008.13318}{{\ttfamily arXiv:2008.13318 [gr-qc]}}.

\bibitem{Kiorpelidi:2022kuo}
S.~Kiorpelidi, G.~Koutsoumbas, A.~Machattou, and E.~Papantonopoulos, ``{Topological black holes with curvature induced scalarization in the extended scalar-tensor theories},'' \href{http://dx.doi.org/10.1103/PhysRevD.105.104039}{{\em Phys. Rev. D} {\bfseries 105} no.~10, (2022) 104039}, \href{http://arxiv.org/abs/2202.00655}{{\ttfamily arXiv:2202.00655 [gr-qc]}}.

\bibitem{Ventagli:2020rnx}
G.~Ventagli, A.~Leh\'ebel, and T.~P. Sotiriou, ``{Onset of spontaneous scalarization in generalized scalar-tensor theories},'' \href{http://dx.doi.org/10.1103/PhysRevD.102.024050}{{\em Phys. Rev. D} {\bfseries 102} no.~2, (2020) 024050}, \href{http://arxiv.org/abs/2006.01153}{{\ttfamily arXiv:2006.01153 [gr-qc]}}.

\bibitem{Thaalba:2022bnt}
F.~Thaalba, G.~Antoniou, and T.~P. Sotiriou, ``{Black hole minimum size and scalar charge in shift-symmetric theories},'' \href{http://dx.doi.org/10.1088/1361-6382/acdd42}{{\em Class. Quant. Grav.} {\bfseries 40} no.~15, (2023) 155002}, \href{http://arxiv.org/abs/2211.05099}{{\ttfamily arXiv:2211.05099 [gr-qc]}}.

\bibitem{Antoniou:2022agj}
G.~Antoniou, C.~F.~B. Macedo, R.~McManus, and T.~P. Sotiriou, ``{Stable spontaneously-scalarized black holes in generalized scalar-tensor theories},'' \href{http://dx.doi.org/10.1103/PhysRevD.106.024029}{{\em Phys. Rev. D} {\bfseries 106} no.~2, (2022) 024029}, \href{http://arxiv.org/abs/2204.01684}{{\ttfamily arXiv:2204.01684 [gr-qc]}}.

\bibitem{Antoniou:2021zoy}
G.~Antoniou, A.~Leh\'ebel, G.~Ventagli, and T.~P. Sotiriou, ``{Black hole scalarization with Gauss-Bonnet and Ricci scalar couplings},'' \href{http://dx.doi.org/10.1103/PhysRevD.104.044002}{{\em Phys. Rev. D} {\bfseries 104} no.~4, (2021) 044002}, \href{http://arxiv.org/abs/2105.04479}{{\ttfamily arXiv:2105.04479 [gr-qc]}}.

\bibitem{Antoniou:2020nax}
G.~Antoniou, L.~Bordin, and T.~P. Sotiriou, ``{Compact object scalarization with general relativity as a cosmic attractor},'' \href{http://dx.doi.org/10.1103/PhysRevD.103.024012}{{\em Phys. Rev. D} {\bfseries 103} no.~2, (2021) 024012}, \href{http://arxiv.org/abs/2004.14985}{{\ttfamily arXiv:2004.14985 [gr-qc]}}.

\bibitem{Andreou:2019ikc}
N.~Andreou, N.~Franchini, G.~Ventagli, and T.~P. Sotiriou, ``{Spontaneous scalarization in generalised scalar-tensor theory},'' \href{http://dx.doi.org/10.1103/PhysRevD.99.124022}{{\em Phys. Rev. D} {\bfseries 99} no.~12, (2019) 124022}, \href{http://arxiv.org/abs/1904.06365}{{\ttfamily arXiv:1904.06365 [gr-qc]}}. [Erratum: Phys.Rev.D 101, 109903 (2020)].

\bibitem{Guajardo:2024hrl}
L.~Guajardo and J.~Oliva, ``{Primary scalar hair in Gauss\textendash{}Bonnet black holes with Thurston horizons},'' \href{http://dx.doi.org/10.1140/epjc/s10052-025-13869-9}{{\em Eur. Phys. J. C} {\bfseries 85} no.~2, (2025) 139}, \href{http://arxiv.org/abs/2412.20134}{{\ttfamily arXiv:2412.20134 [hep-th]}}.

\bibitem{Charmousis:2025jpx}
C.~Charmousis, P.~G.~S. Fernandes, and M.~Hassaine, ``{Proca theory of four-dimensional regularized Gauss-Bonnet gravity and black holes with primary hair},'' \href{http://dx.doi.org/10.1103/9f2w-3kly}{{\em Phys. Rev. D} {\bfseries 111} no.~12, (2025) 124008}, \href{http://arxiv.org/abs/2504.13084}{{\ttfamily arXiv:2504.13084 [gr-qc]}}.

\bibitem{Bakopoulos:2021liw}
A.~Bakopoulos, C.~Charmousis, and P.~Kanti, ``{Traversable wormholes in beyond Horndeski theories},'' \href{http://dx.doi.org/10.1088/1475-7516/2022/05/022}{{\em JCAP} {\bfseries 05} no.~05, (2022) 022}, \href{http://arxiv.org/abs/2111.09857}{{\ttfamily arXiv:2111.09857 [gr-qc]}}.

\bibitem{Bakopoulos:2023tso}
A.~Bakopoulos, N.~Chatzifotis, C.~Erices, and E.~Papantonopoulos, ``{Stealth Ellis wormholes in Horndeski theories},'' \href{http://dx.doi.org/10.1088/1475-7516/2023/11/055}{{\em JCAP} {\bfseries 11} (2023) 055}, \href{http://arxiv.org/abs/2306.16768}{{\ttfamily arXiv:2306.16768 [hep-th]}}.

\bibitem{Antoniou:2019awm}
G.~Antoniou, A.~Bakopoulos, P.~Kanti, B.~Kleihaus, and J.~Kunz, ``{Novel Einstein\textendash{}scalar-Gauss-Bonnet wormholes without exotic matter},'' \href{http://dx.doi.org/10.1103/PhysRevD.101.024033}{{\em Phys. Rev. D} {\bfseries 101} no.~2, (2020) 024033}, \href{http://arxiv.org/abs/1904.13091}{{\ttfamily arXiv:1904.13091 [hep-th]}}.

\bibitem{Kanti:2011jz}
P.~Kanti, B.~Kleihaus, and J.~Kunz, ``{Wormholes in Dilatonic Einstein-Gauss-Bonnet Theory},'' \href{http://dx.doi.org/10.1103/PhysRevLett.107.271101}{{\em Phys. Rev. Lett.} {\bfseries 107} (2011) 271101}, \href{http://arxiv.org/abs/1108.3003}{{\ttfamily arXiv:1108.3003 [gr-qc]}}.

\bibitem{Kanti:2011yv}
P.~Kanti, B.~Kleihaus, and J.~Kunz, ``{Stable Lorentzian Wormholes in Dilatonic Einstein-Gauss-Bonnet Theory},'' \href{http://dx.doi.org/10.1103/PhysRevD.85.044007}{{\em Phys. Rev. D} {\bfseries 85} (2012) 044007}, \href{http://arxiv.org/abs/1111.4049}{{\ttfamily arXiv:1111.4049 [hep-th]}}.

\bibitem{Karakasis:2021tqx}
T.~Karakasis, E.~Papantonopoulos, and C.~Vlachos, ``{f(R) gravity wormholes sourced by a phantom scalar field},'' \href{http://dx.doi.org/10.1103/PhysRevD.105.024006}{{\em Phys. Rev. D} {\bfseries 105} no.~2, (2022) 024006}, \href{http://arxiv.org/abs/2107.09713}{{\ttfamily arXiv:2107.09713 [gr-qc]}}.

\bibitem{Chew:2016epf}
X.~Y. Chew, B.~Kleihaus, and J.~Kunz, ``{Geometry of Spinning Ellis Wormholes},'' \href{http://dx.doi.org/10.1103/PhysRevD.94.104031}{{\em Phys. Rev. D} {\bfseries 94} no.~10, (2016) 104031}, \href{http://arxiv.org/abs/1608.05253}{{\ttfamily arXiv:1608.05253 [gr-qc]}}.

\bibitem{Chew:2018vjp}
X.~Y. Chew, B.~Kleihaus, and J.~Kunz, ``{Spinning Wormholes in Scalar-Tensor Theory},'' \href{http://dx.doi.org/10.1103/PhysRevD.97.064026}{{\em Phys. Rev. D} {\bfseries 97} no.~6, (2018) 064026}, \href{http://arxiv.org/abs/1802.00365}{{\ttfamily arXiv:1802.00365 [gr-qc]}}.

\bibitem{Blazquez-Salcedo:2018ipc}
J.~L. Bl\'azquez-Salcedo, X.~Y. Chew, and J.~Kunz, ``{Scalar and axial quasinormal modes of massive static phantom wormholes},'' \href{http://dx.doi.org/10.1103/PhysRevD.98.044035}{{\em Phys. Rev. D} {\bfseries 98} no.~4, (2018) 044035}, \href{http://arxiv.org/abs/1806.03282}{{\ttfamily arXiv:1806.03282 [gr-qc]}}.

\bibitem{Chew:2019lsa}
X.~Y. Chew, V.~Dzhunushaliev, V.~Folomeev, B.~Kleihaus, and J.~Kunz, ``{Rotating wormhole solutions with a complex phantom scalar field},'' \href{http://dx.doi.org/10.1103/PhysRevD.100.044019}{{\em Phys. Rev. D} {\bfseries 100} no.~4, (2019) 044019}, \href{http://arxiv.org/abs/1906.08742}{{\ttfamily arXiv:1906.08742 [gr-qc]}}.

\bibitem{Chew:2020lkj}
X.~Y. Chew, G.~Tumurtushaa, and D.-h. Yeom, ``{Euclidean wormholes in Gauss\textendash{}Bonnet-dilaton gravity},'' \href{http://dx.doi.org/10.1016/j.dark.2021.100811}{{\em Phys. Dark Univ.} {\bfseries 32} (2021) 100811}, \href{http://arxiv.org/abs/2006.04344}{{\ttfamily arXiv:2006.04344 [gr-qc]}}.

\bibitem{Blazquez-Salcedo:2020nsa}
J.~L. Bl\'azquez-Salcedo, X.~Y. Chew, J.~Kunz, and D.-H. Yeom, ``{Ellis wormholes in anti-de Sitter space},'' \href{http://dx.doi.org/10.1140/epjc/s10052-021-09645-0}{{\em Eur. Phys. J. C} {\bfseries 81} no.~9, (2021) 858}, \href{http://arxiv.org/abs/2012.06213}{{\ttfamily arXiv:2012.06213 [gr-qc]}}.

\bibitem{Kleihaus:2019rbg}
B.~Kleihaus, J.~Kunz, and P.~Kanti, ``{Particle-like ultracompact objects in Einstein-scalar-Gauss-Bonnet theories},'' \href{http://dx.doi.org/10.1016/j.physletb.2020.135401}{{\em Phys. Lett. B} {\bfseries 804} (2020) 135401}, \href{http://arxiv.org/abs/1910.02121}{{\ttfamily arXiv:1910.02121 [gr-qc]}}.

\bibitem{Kleihaus:2020qwo}
B.~Kleihaus, J.~Kunz, and P.~Kanti, ``{Properties of ultracompact particlelike solutions in Einstein-scalar-Gauss-Bonnet theories},'' \href{http://dx.doi.org/10.1103/PhysRevD.102.024070}{{\em Phys. Rev. D} {\bfseries 102} no.~2, (2020) 024070}, \href{http://arxiv.org/abs/2005.07650}{{\ttfamily arXiv:2005.07650 [gr-qc]}}.

\bibitem{Bekenstein:1972tm}
J.~D. Bekenstein, ``{Black holes and the second law},'' \href{http://dx.doi.org/10.1007/BF02757029}{{\em Lett. Nuovo Cim.} {\bfseries 4} (1972) 737--740}.

\bibitem{Bekenstein:1973ur}
J.~D. Bekenstein, ``{Black holes and entropy},'' \href{http://dx.doi.org/10.1103/PhysRevD.7.2333}{{\em Phys. Rev. D} {\bfseries 7} (1973) 2333--2346}.

\bibitem{Hawking:1976de}
S.~W. Hawking, ``{Black Holes and Thermodynamics},'' \href{http://dx.doi.org/10.1103/PhysRevD.13.191}{{\em Phys. Rev. D} {\bfseries 13} (1976) 191--197}.

\bibitem{Hawking:1975vcx}
S.~W. Hawking, ``{Particle Creation by Black Holes},'' \href{http://dx.doi.org/10.1007/BF02345020}{{\em Commun. Math. Phys.} {\bfseries 43} (1975) 199--220}. [Erratum: Commun.Math.Phys. 46, 206 (1976)].

\bibitem{Bardeen:1973gs}
J.~M. Bardeen, B.~Carter, and S.~W. Hawking, ``{The Four laws of black hole mechanics},'' \href{http://dx.doi.org/10.1007/BF01645742}{{\em Commun. Math. Phys.} {\bfseries 31} (1973) 161--170}.

\bibitem{Carlip:2014pma}
S.~Carlip, ``{Black Hole Thermodynamics},'' \href{http://dx.doi.org/10.1142/S0218271814300237}{{\em Int. J. Mod. Phys. D} {\bfseries 23} (2014) 1430023}, \href{http://arxiv.org/abs/1410.1486}{{\ttfamily arXiv:1410.1486 [gr-qc]}}.

\bibitem{Giddings:1995gd}
S.~B. Giddings, ``{The Black hole information paradox},'' in {\em {PASCOS / HOPKINS 1995 (Joint Meeting of the International Symposium on Particles, Strings and Cosmology and the 19th Johns Hopkins Workshop on Current Problems in Particle Theory)}}, pp.~415--428.
\newblock 8, 1995.
\newblock \href{http://arxiv.org/abs/hep-th/9508151}{{\ttfamily arXiv:hep-th/9508151}}.

\bibitem{Mathur:2008wi}
S.~D. Mathur, ``{What Exactly is the Information Paradox?},'' \href{http://dx.doi.org/10.1007/978-3-540-88460-6_1}{{\em Lect. Notes Phys.} {\bfseries 769} (2009) 3--48}, \href{http://arxiv.org/abs/0803.2030}{{\ttfamily arXiv:0803.2030 [hep-th]}}.

\bibitem{Hawking:1971tu}
S.~W. Hawking, ``{Gravitational radiation from colliding black holes},'' \href{http://dx.doi.org/10.1103/PhysRevLett.26.1344}{{\em Phys. Rev. Lett.} {\bfseries 26} (1971) 1344--1346}.

\bibitem{Brustein:2007jj}
R.~Brustein, D.~Gorbonos, and M.~Hadad, ``{Wald's entropy is equal to a quarter of the horizon area in units of the effective gravitational coupling},'' \href{http://dx.doi.org/10.1103/PhysRevD.79.044025}{{\em Phys. Rev. D} {\bfseries 79} (2009) 044025}, \href{http://arxiv.org/abs/0712.3206}{{\ttfamily arXiv:0712.3206 [hep-th]}}.

\bibitem{Martinez:1996gn}
C.~Martinez and J.~Zanelli, ``{Conformally dressed black hole in (2+1)-dimensions},'' \href{http://dx.doi.org/10.1103/PhysRevD.54.3830}{{\em Phys. Rev. D} {\bfseries 54} (1996) 3830--3833}, \href{http://arxiv.org/abs/gr-qc/9604021}{{\ttfamily arXiv:gr-qc/9604021}}.

\bibitem{Martinez:2005di}
C.~Martinez, J.~P. Staforelli, and R.~Troncoso, ``{Topological black holes dressed with a conformally coupled scalar field and electric charge},'' \href{http://dx.doi.org/10.1103/PhysRevD.74.044028}{{\em Phys. Rev. D} {\bfseries 74} (2006) 044028}, \href{http://arxiv.org/abs/hep-th/0512022}{{\ttfamily arXiv:hep-th/0512022}}.

\bibitem{Barlow:2005yd}
A.-M. Barlow, D.~Doherty, and E.~Winstanley, ``{Thermodynamics of de Sitter black holes with a conformally coupled scalar field},'' \href{http://dx.doi.org/10.1103/PhysRevD.72.024008}{{\em Phys. Rev. D} {\bfseries 72} (2005) 024008}, \href{http://arxiv.org/abs/gr-qc/0504087}{{\ttfamily arXiv:gr-qc/0504087}}.

\bibitem{Anastasiou:2022wjq}
G.~Anastasiou, I.~J. Araya, M.~Busnego-Barrientos, C.~Corral, and N.~Merino, ``{Conformal renormalization of scalar-tensor theories},'' \href{http://dx.doi.org/10.1103/PhysRevD.107.104049}{{\em Phys. Rev. D} {\bfseries 107} no.~10, (2023) 104049}, \href{http://arxiv.org/abs/2212.04364}{{\ttfamily arXiv:2212.04364 [hep-th]}}.

\bibitem{Bravo-Gaete:2025vyd}
M.~Bravo-Gaete, F.~F. Santos, J.~A. Herrera-Mendoza, and D.~F. Higuita-Borja, ``{Rotating axionic AdS$_4$ black hole dressed with a scalar field},'' \href{http://arxiv.org/abs/2504.17081}{{\ttfamily arXiv:2504.17081 [hep-th]}}.

\bibitem{Sebastiani:2010kv}
L.~Sebastiani and S.~Zerbini, ``{Static Spherically Symmetric Solutions in F(R) Gravity},'' \href{http://dx.doi.org/10.1140/epjc/s10052-011-1591-8}{{\em Eur. Phys. J. C} {\bfseries 71} (2011) 1591}, \href{http://arxiv.org/abs/1012.5230}{{\ttfamily arXiv:1012.5230 [gr-qc]}}.

\bibitem{Faraoni:2010yi}
V.~Faraoni, ``{Black hole entropy in scalar-tensor and f(R) gravity: An Overview},'' \href{http://dx.doi.org/10.3390/e12051246}{{\em Entropy} {\bfseries 12} (2010) 1246}, \href{http://arxiv.org/abs/1005.2327}{{\ttfamily arXiv:1005.2327 [gr-qc]}}.

\bibitem{Cai:1996pj}
R.-G. Cai and Y.~S. Myung, ``{Black holes in the Brans-Dicke-Maxwell theory},'' \href{http://dx.doi.org/10.1103/PhysRevD.56.3466}{{\em Phys. Rev. D} {\bfseries 56} (1997) 3466--3470}, \href{http://arxiv.org/abs/gr-qc/9702037}{{\ttfamily arXiv:gr-qc/9702037}}.

\bibitem{Chew:2022enh}
X.~Y. Chew, D.-h. Yeom, and J.~L. Bl\'azquez-Salcedo, ``{Properties of scalar hairy black holes and scalarons with asymmetric potential},'' \href{http://dx.doi.org/10.1103/PhysRevD.108.044020}{{\em Phys. Rev. D} {\bfseries 108} no.~4, (2023) 044020}, \href{http://arxiv.org/abs/2210.01313}{{\ttfamily arXiv:2210.01313 [gr-qc]}}.

\bibitem{Chew:2023olq}
X.~Y. Chew and K.-G. Lim, ``{Scalar hairy black holes with an inverted Mexican-hat potential},'' \href{http://dx.doi.org/10.1103/PhysRevD.109.064039}{{\em Phys. Rev. D} {\bfseries 109} no.~6, (2024) 064039}, \href{http://arxiv.org/abs/2307.13972}{{\ttfamily arXiv:2307.13972 [gr-qc]}}.

\bibitem{Chew:2024rin}
X.~Y. Chew and D.-h. Yeom, ``{Hairy Reissner-Nordstr\"om black holes with asymmetric vacua},'' \href{http://dx.doi.org/10.1103/PhysRevD.110.044036}{{\em Phys. Rev. D} {\bfseries 110} no.~4, (2024) 044036}, \href{http://arxiv.org/abs/2401.09039}{{\ttfamily arXiv:2401.09039 [gr-qc]}}.

\bibitem{Chew:2024evh}
X.~Y. Chew and Y.~S. Myung, ``{Simplest model of a scalarized black hole in the Einstein-Klein-Gordon theory},'' \href{http://dx.doi.org/10.1103/PhysRevD.110.044011}{{\em Phys. Rev. D} {\bfseries 110} no.~4, (2024) 044011}, \href{http://arxiv.org/abs/2405.04921}{{\ttfamily arXiv:2405.04921 [gr-qc]}}.

\bibitem{Karakasis:2022fep}
T.~Karakasis, E.~Papantonopoulos, Z.-Y. Tang, and B.~Wang, ``{Rotating (2+1)-dimensional black holes in Einstein-Maxwell-dilaton theory},'' \href{http://dx.doi.org/10.1103/PhysRevD.107.024043}{{\em Phys. Rev. D} {\bfseries 107} no.~2, (2023) 024043}, \href{http://arxiv.org/abs/2210.15704}{{\ttfamily arXiv:2210.15704 [gr-qc]}}.

\bibitem{Gonzalez:2013aca}
P.~A. Gonz\'alez, E.~Papantonopoulos, J.~Saavedra, and Y.~V\'asquez, ``{Four-Dimensional Asymptotically AdS Black Holes with Scalar Hair},'' \href{http://dx.doi.org/10.1007/JHEP12(2013)021}{{\em JHEP} {\bfseries 12} (2013) 021}, \href{http://arxiv.org/abs/1309.2161}{{\ttfamily arXiv:1309.2161 [gr-qc]}}.

\bibitem{Karakasis:2022xzm}
T.~Karakasis, G.~Koutsoumbas, A.~Machattou, and E.~Papantonopoulos, ``{Magnetically charged Euler-Heisenberg black holes with scalar hair},'' \href{http://dx.doi.org/10.1103/PhysRevD.106.104006}{{\em Phys. Rev. D} {\bfseries 106} no.~10, (2022) 104006}, \href{http://arxiv.org/abs/2207.13146}{{\ttfamily arXiv:2207.13146 [gr-qc]}}.

\bibitem{Cai:2001dz}
R.-G. Cai, ``{Gauss-Bonnet black holes in AdS spaces},'' \href{http://dx.doi.org/10.1103/PhysRevD.65.084014}{{\em Phys. Rev. D} {\bfseries 65} (2002) 084014}, \href{http://arxiv.org/abs/hep-th/0109133}{{\ttfamily arXiv:hep-th/0109133}}.

\bibitem{Jacobson:1993vj}
T.~Jacobson, G.~Kang, and R.~C. Myers, ``{On black hole entropy},'' \href{http://dx.doi.org/10.1103/PhysRevD.49.6587}{{\em Phys. Rev. D} {\bfseries 49} (1994) 6587--6598}, \href{http://arxiv.org/abs/gr-qc/9312023}{{\ttfamily arXiv:gr-qc/9312023}}.

\bibitem{Tachikawa:2006sz}
Y.~Tachikawa, ``{Black hole entropy in the presence of Chern-Simons terms},'' \href{http://dx.doi.org/10.1088/0264-9381/24/3/014}{{\em Class. Quant. Grav.} {\bfseries 24} (2007) 737--744}, \href{http://arxiv.org/abs/hep-th/0611141}{{\ttfamily arXiv:hep-th/0611141}}.

\bibitem{Feng:2015oea}
X.-H. Feng, H.-S. Liu, H.~L\"u, and C.~N. Pope, ``{Black Hole Entropy and Viscosity Bound in Horndeski Gravity},'' \href{http://dx.doi.org/10.1007/JHEP11(2015)176}{{\em JHEP} {\bfseries 11} (2015) 176}, \href{http://arxiv.org/abs/1509.07142}{{\ttfamily arXiv:1509.07142 [hep-th]}}.

\bibitem{Bravo-Gaete:2014haa}
M.~Bravo-Gaete and M.~Hassaine, ``{Thermodynamics of a BTZ black hole solution with an Horndeski source},'' \href{http://dx.doi.org/10.1103/PhysRevD.90.024008}{{\em Phys. Rev. D} {\bfseries 90} no.~2, (2014) 024008}, \href{http://arxiv.org/abs/1405.4935}{{\ttfamily arXiv:1405.4935 [hep-th]}}.

\bibitem{Bravo-Gaete:2021hlc}
M.~Bravo-Gaete and M.~M. Stetsko, ``{Planar black holes configurations and shear viscosity in arbitrary dimensions with shift and reflection symmetric scalar-tensor theories},'' \href{http://dx.doi.org/10.1103/PhysRevD.105.024038}{{\em Phys. Rev. D} {\bfseries 105} no.~2, (2022) 024038}, \href{http://arxiv.org/abs/2111.10925}{{\ttfamily arXiv:2111.10925 [hep-th]}}.

\bibitem{Bravo-Gaete:2021hza}
M.~Bravo-Gaete, C.~G. Gaete, L.~Guajardo, and S.~G. Rodr\'\i{}guez, ``{Towards the emergence of nonzero thermodynamical quantities for Lanczos-Lovelock black holes dressed with a scalar field},'' \href{http://dx.doi.org/10.1103/PhysRevD.104.044027}{{\em Phys. Rev. D} {\bfseries 104} no.~4, (2021) 044027}, \href{http://arxiv.org/abs/2103.15634}{{\ttfamily arXiv:2103.15634 [gr-qc]}}.

\bibitem{Cisterna:2025vxk}
A.~Cisterna, M.~Hassaine, and U.~Hernandez-Vera, ``{Thermodynamics of four-dimensional regular black holes with an infinite tower of regularized curvature corrections},'' \href{http://dx.doi.org/10.1103/6f3b-8794}{{\em Phys. Rev. D} {\bfseries 112} no.~6, (2025) 064036}, \href{http://arxiv.org/abs/2505.23467}{{\ttfamily arXiv:2505.23467 [gr-qc]}}.

\bibitem{Hennigar:2025ftm}
R.~A. Hennigar, D.~Kubiz{\v{n}}{\'a}k, S.~Murk, and I.~Soranidis, ``{Thermodynamics of Regular Black Holes in Anti-de Sitter Space},'' \href{http://arxiv.org/abs/2505.11623}{{\ttfamily arXiv:2505.11623 [gr-qc]}}.

\bibitem{Bueno:2025zaj}
P.~Bueno, P.~A. Cano, R.~A. Hennigar, and {\'A}.~J. Murcia, ``{Regular black hole formation in four-dimensional non-polynomial gravities},'' \href{http://arxiv.org/abs/2509.19016}{{\ttfamily arXiv:2509.19016 [gr-qc]}}.

\bibitem{Bueno:2024dgm}
P.~Bueno, P.~A. Cano, and R.~A. Hennigar, ``{Regular black holes from pure gravity},'' \href{http://dx.doi.org/10.1016/j.physletb.2025.139260}{{\em Phys. Lett. B} {\bfseries 861} (2025) 139260}, \href{http://arxiv.org/abs/2403.04827}{{\ttfamily arXiv:2403.04827 [gr-qc]}}.

\bibitem{Fernandes:2025eoc}
P.~G.~S. Fernandes, ``{Regular BTZ black holes from an infinite tower of corrections},'' \href{http://dx.doi.org/10.1016/j.physletb.2025.139772}{{\em Phys. Lett. B} {\bfseries 868} (2025) 139772}, \href{http://arxiv.org/abs/2504.08565}{{\ttfamily arXiv:2504.08565 [gr-qc]}}.

\bibitem{Fernandes:2025fnz}
P.~G.~S. Fernandes, ``{Singularity resolution and inflation from an infinite tower of regularized curvature corrections},'' \href{http://arxiv.org/abs/2504.07692}{{\ttfamily arXiv:2504.07692 [gr-qc]}}.

\bibitem{Minamitsuji:2023nvh}
M.~Minamitsuji and K.-i. Maeda, ``{Black hole thermodynamics in Horndeski theories},'' \href{http://dx.doi.org/10.1103/PhysRevD.108.084061}{{\em Phys. Rev. D} {\bfseries 108} no.~8, (2023) 084061}, \href{http://arxiv.org/abs/2308.01082}{{\ttfamily arXiv:2308.01082 [gr-qc]}}.

\bibitem{Wald:1993nt}
R.~M. Wald, ``{Black hole entropy is the Noether charge},'' \href{http://dx.doi.org/10.1103/PhysRevD.48.R3427}{{\em Phys. Rev. D} {\bfseries 48} no.~8, (1993) R3427--R3431}, \href{http://arxiv.org/abs/gr-qc/9307038}{{\ttfamily arXiv:gr-qc/9307038}}.

\bibitem{Iyer:1994ys}
V.~Iyer and R.~M. Wald, ``{Some properties of Noether charge and a proposal for dynamical black hole entropy},'' \href{http://dx.doi.org/10.1103/PhysRevD.50.846}{{\em Phys. Rev. D} {\bfseries 50} (1994) 846--864}, \href{http://arxiv.org/abs/gr-qc/9403028}{{\ttfamily arXiv:gr-qc/9403028}}.

\bibitem{Bakopoulos:2024zke}
A.~Bakopoulos, T.~Karakasis, and E.~Papantonopoulos, ``{Thermodynamics of stealth black holes},'' \href{http://dx.doi.org/10.1103/PhysRevD.111.024065}{{\em Phys. Rev. D} {\bfseries 111} no.~2, (2025) 024065}, \href{http://arxiv.org/abs/2410.14451}{{\ttfamily arXiv:2410.14451 [hep-th]}}.

\bibitem{Hajian:2020dcq}
K.~Hajian, S.~Liberati, M.~M. Sheikh-Jabbari, and M.~H. Vahidinia, ``{On Black Hole Temperature in Horndeski Gravity},'' \href{http://dx.doi.org/10.1016/j.physletb.2020.136002}{{\em Phys. Lett. B} {\bfseries 812} (2021) 136002}, \href{http://arxiv.org/abs/2005.12985}{{\ttfamily arXiv:2005.12985 [gr-qc]}}.

\bibitem{Crisostomi:2016tcp}
M.~Crisostomi, M.~Hull, K.~Koyama, and G.~Tasinato, ``{Horndeski: beyond, or not beyond?},'' \href{http://dx.doi.org/10.1088/1475-7516/2016/03/038}{{\em JCAP} {\bfseries 03} (2016) 038}, \href{http://arxiv.org/abs/1601.04658}{{\ttfamily arXiv:1601.04658 [hep-th]}}.

\bibitem{Bakopoulos:2024ogt}
A.~Bakopoulos, N.~Chatzifotis, and T.~Karakasis, ``{Thermodynamics of black holes featuring primary scalar hair},'' \href{http://dx.doi.org/10.1103/PhysRevD.110.L101502}{{\em Phys. Rev. D} {\bfseries 110} no.~10, (2024) L101502}, \href{http://arxiv.org/abs/2404.07522}{{\ttfamily arXiv:2404.07522 [hep-th]}}.

\bibitem{Bakopoulos:2024hah}
A.~Bakopoulos, T.~Karakasis, N.~E. Mavromatos, T.~Nakas, and E.~Papantonopoulos, ``{Exact black holes in string-inspired Euler-Heisenberg theory},'' \href{http://dx.doi.org/10.1103/PhysRevD.110.024014}{{\em Phys. Rev. D} {\bfseries 110} no.~2, (2024) 024014}, \href{http://arxiv.org/abs/2402.12459}{{\ttfamily arXiv:2402.12459 [hep-th]}}.

\bibitem{Gibbons:1976ue}
G.~W. Gibbons and S.~W. Hawking, ``{Action Integrals and Partition Functions in Quantum Gravity},'' \href{http://dx.doi.org/10.1103/PhysRevD.15.2752}{{\em Phys. Rev. D} {\bfseries 15} (1977) 2752--2756}.

\bibitem{Witten:2024upt}
E.~Witten, ``{Introduction to Black Hole Thermodynamics},'' \href{http://arxiv.org/abs/2412.16795}{{\ttfamily arXiv:2412.16795 [hep-th]}}.

\bibitem{Rinaldi:2012vy}
M.~Rinaldi, ``{Black holes with non-minimal derivative coupling},'' \href{http://dx.doi.org/10.1103/PhysRevD.86.084048}{{\em Phys. Rev. D} {\bfseries 86} (2012) 084048}, \href{http://arxiv.org/abs/1208.0103}{{\ttfamily arXiv:1208.0103 [gr-qc]}}.

\bibitem{Hassaine:2007py}
M.~Hassaine and C.~Martinez, ``{Higher-dimensional black holes with a conformally invariant Maxwell source},'' \href{http://dx.doi.org/10.1103/PhysRevD.75.027502}{{\em Phys. Rev. D} {\bfseries 75} (2007) 027502}, \href{http://arxiv.org/abs/hep-th/0701058}{{\ttfamily arXiv:hep-th/0701058}}.

\bibitem{Gonzalez:2009nn}
H.~A. Gonzalez, M.~Hassaine, and C.~Martinez, ``{Thermodynamics of charged black holes with a nonlinear electrodynamics source},'' \href{http://dx.doi.org/10.1103/PhysRevD.80.104008}{{\em Phys. Rev. D} {\bfseries 80} (2009) 104008}, \href{http://arxiv.org/abs/0909.1365}{{\ttfamily arXiv:0909.1365 [hep-th]}}.

\bibitem{Babichev:2017guv}
E.~Babichev, C.~Charmousis, and A.~Leh\'ebel, ``{Asymptotically flat black holes in Horndeski theory and beyond},'' \href{http://dx.doi.org/10.1088/1475-7516/2017/04/027}{{\em JCAP} {\bfseries 04} (2017) 027}, \href{http://arxiv.org/abs/1702.01938}{{\ttfamily arXiv:1702.01938 [gr-qc]}}.

\bibitem{Jacobson:2007tj}
T.~Jacobson, ``{When is g(tt) g(rr) = -1?},'' \href{http://dx.doi.org/10.1088/0264-9381/24/22/N02}{{\em Class. Quant. Grav.} {\bfseries 24} (2007) 5717--5719}, \href{http://arxiv.org/abs/0707.3222}{{\ttfamily arXiv:0707.3222 [gr-qc]}}.

\bibitem{Dorlis:2023qug}
P.~Dorlis, N.~E. Mavromatos, and S.-N. Vlachos, ``{Bypassing Bekenstein\textquoteright{}s no-scalar-hair theorem without violating the energy conditions},'' \href{http://dx.doi.org/10.1103/PhysRevD.108.064004}{{\em Phys. Rev. D} {\bfseries 108} no.~6, (2023) 064004}, \href{http://arxiv.org/abs/2305.18031}{{\ttfamily arXiv:2305.18031 [gr-qc]}}.

\bibitem{Neupane:2002bf}
I.~P. Neupane, ``{Black hole entropy in string generated gravity models},'' \href{http://dx.doi.org/10.1103/PhysRevD.67.061501}{{\em Phys. Rev. D} {\bfseries 67} (2003) 061501}, \href{http://arxiv.org/abs/hep-th/0212092}{{\ttfamily arXiv:hep-th/0212092}}.

\bibitem{Bergliaffa:2021diw}
S.~E.~P. Bergliaffa, R.~Maier, and N.~d.~O. Silvano, ``{Hairy Black Holes from Horndeski Theory},'' \href{http://arxiv.org/abs/2107.07839}{{\ttfamily arXiv:2107.07839 [gr-qc]}}.

\bibitem{Yang:2024cjf}
Z.-H. Yang, Y.-H. Lei, X.-M. Kuang, and B.~Wang, ``{Gravitational odd-parity perturbation of a Horndeski hairy black hole: quasinormal mode and parameter constraint},'' \href{http://arxiv.org/abs/2408.07418}{{\ttfamily arXiv:2408.07418 [gr-qc]}}.

\bibitem{Jha:2022tdl}
S.~K. Jha, M.~Khodadi, A.~Rahaman, and A.~Sheykhi, ``{Superradiant energy extraction from rotating hairy Horndeski black holes},'' \href{http://dx.doi.org/10.1103/PhysRevD.107.084052}{{\em Phys. Rev. D} {\bfseries 107} no.~8, (2023) 084052}, \href{http://arxiv.org/abs/2212.13051}{{\ttfamily arXiv:2212.13051 [gr-qc]}}.

\bibitem{Gohain:2025jbz}
M.~M. Gohain and K.~Bhuyan, ``{Dark Matter Surrounded Quartic Square-root Horndeski Black Hole: Thermodynamics, Optical Properties and Quasinormal Oscillations},'' \href{http://arxiv.org/abs/2506.15763}{{\ttfamily arXiv:2506.15763 [gr-qc]}}.

\bibitem{Teitelboim:1994az}
C.~Teitelboim, ``{Action and entropy of extreme and nonextreme black holes},'' \href{http://dx.doi.org/10.1103/PhysRevD.52.6201}{{\em Phys. Rev. D} {\bfseries 51} (1995) 4315}, \href{http://arxiv.org/abs/hep-th/9410103}{{\ttfamily arXiv:hep-th/9410103}}. [Erratum: Phys.Rev.D 52, 6201 (1995)].

\bibitem{Hawking:1994ii}
S.~W. Hawking, G.~T. Horowitz, and S.~F. Ross, ``{Entropy, Area, and black hole pairs},'' \href{http://dx.doi.org/10.1103/PhysRevD.51.4302}{{\em Phys. Rev. D} {\bfseries 51} (1995) 4302--4314}, \href{http://arxiv.org/abs/gr-qc/9409013}{{\ttfamily arXiv:gr-qc/9409013}}.

\bibitem{Strominger:1996sh}
A.~Strominger and C.~Vafa, ``{Microscopic origin of the Bekenstein-Hawking entropy},'' \href{http://dx.doi.org/10.1016/0370-2693(96)00345-0}{{\em Phys. Lett. B} {\bfseries 379} (1996) 99--104}, \href{http://arxiv.org/abs/hep-th/9601029}{{\ttfamily arXiv:hep-th/9601029}}.

\bibitem{Guajardo:2023uix}
L.~Guajardo, ``{Black string spectrum of shift-symmetric Horndeski theories},'' \href{http://dx.doi.org/10.1103/PhysRevD.108.124041}{{\em Phys. Rev. D} {\bfseries 108} no.~12, (2023) 124041}, \href{http://arxiv.org/abs/2304.14240}{{\ttfamily arXiv:2304.14240 [hep-th]}}.

\bibitem{Charmousis:2011bf}
C.~Charmousis, E.~J. Copeland, A.~Padilla, and P.~M. Saffin, ``{General second order scalar-tensor theory, self tuning, and the Fab Four},'' \href{http://dx.doi.org/10.1103/PhysRevLett.108.051101}{{\em Phys. Rev. Lett.} {\bfseries 108} (2012) 051101}, \href{http://arxiv.org/abs/1106.2000}{{\ttfamily arXiv:1106.2000 [hep-th]}}.

\bibitem{Charmousis:2011ea}
C.~Charmousis, E.~J. Copeland, A.~Padilla, and P.~M. Saffin, ``{Self-tuning and the derivation of a class of scalar-tensor theories},'' \href{http://dx.doi.org/10.1103/PhysRevD.85.104040}{{\em Phys. Rev. D} {\bfseries 85} (2012) 104040}, \href{http://arxiv.org/abs/1112.4866}{{\ttfamily arXiv:1112.4866 [hep-th]}}.

\bibitem{Correa:2013bza}
F.~Correa and M.~Hassaine, ``{Thermodynamics of Lovelock black holes with a nonminimal scalar field},'' \href{http://dx.doi.org/10.1007/JHEP02(2014)014}{{\em JHEP} {\bfseries 02} (2014) 014}, \href{http://arxiv.org/abs/1312.4516}{{\ttfamily arXiv:1312.4516 [hep-th]}}.

\bibitem{Minamitsuji:2013ura}
M.~Minamitsuji, ``{Solutions in the scalar-tensor theory with nonminimal derivative coupling},'' \href{http://dx.doi.org/10.1103/PhysRevD.89.064017}{{\em Phys. Rev. D} {\bfseries 89} (2014) 064017}, \href{http://arxiv.org/abs/1312.3759}{{\ttfamily arXiv:1312.3759 [gr-qc]}}.

\bibitem{Arratia:2020hoy}
E.~Arratia, C.~Corral, J.~Figueroa, and L.~Sanhueza, ``{Hairy Taub-NUT/bolt-AdS solutions in Horndeski theory},'' \href{http://dx.doi.org/10.1103/PhysRevD.103.064068}{{\em Phys. Rev. D} {\bfseries 103} no.~6, (2021) 064068}, \href{http://arxiv.org/abs/2010.02460}{{\ttfamily arXiv:2010.02460 [hep-th]}}.

\bibitem{Caceres:2023gfa}
N.~Caceres, C.~Corral, F.~Diaz, and R.~Olea, ``{Holographic renormalization of Horndeski gravity},'' \href{http://dx.doi.org/10.1007/JHEP05(2024)125}{{\em JHEP} {\bfseries 05} (2024) 125}, \href{http://arxiv.org/abs/2311.04054}{{\ttfamily arXiv:2311.04054 [hep-th]}}.

\bibitem{Anabalon:2013oea}
A.~Anabalon, A.~Cisterna, and J.~Oliva, ``{Asymptotically locally AdS and flat black holes in Horndeski theory},'' \href{http://dx.doi.org/10.1103/PhysRevD.89.084050}{{\em Phys. Rev. D} {\bfseries 89} (2014) 084050}, \href{http://arxiv.org/abs/1312.3597}{{\ttfamily arXiv:1312.3597 [gr-qc]}}.

\bibitem{Alkac:2024hvu}
G.~Alkac, L.~Guajardo, and H.~Ozsahin, ``{Microscopic entropy of static black holes in 3D Lovelock gravities},'' \href{http://dx.doi.org/10.1103/PhysRevD.111.044006}{{\em Phys. Rev. D} {\bfseries 111} no.~4, (2025) 044006}, \href{http://arxiv.org/abs/2409.03865}{{\ttfamily arXiv:2409.03865 [hep-th]}}.

\bibitem{Hennigar:2020fkv}
R.~A. Hennigar, D.~Kubiznak, R.~B. Mann, and C.~Pollack, ``{Lower-dimensional Gauss\textendash{}Bonnet gravity and BTZ black holes},'' \href{http://dx.doi.org/10.1016/j.physletb.2020.135657}{{\em Phys. Lett. B} {\bfseries 808} (2020) 135657}, \href{http://arxiv.org/abs/2004.12995}{{\ttfamily arXiv:2004.12995 [gr-qc]}}.

\bibitem{Cardenas:2014kaa}
M.~Cardenas, O.~Fuentealba, and C.~Mart\'\i{}nez, ``{Three-dimensional black holes with conformally coupled scalar and gauge fields},'' \href{http://dx.doi.org/10.1103/PhysRevD.90.124072}{{\em Phys. Rev. D} {\bfseries 90} no.~12, (2014) 124072}, \href{http://arxiv.org/abs/1408.1401}{{\ttfamily arXiv:1408.1401 [hep-th]}}.

\bibitem{Priyadarshinee:2023cmi}
S.~Priyadarshinee and S.~Mahapatra, ``{Analytic three-dimensional primary hair charged black holes and thermodynamics},'' \href{http://dx.doi.org/10.1103/PhysRevD.108.044017}{{\em Phys. Rev. D} {\bfseries 108} no.~4, (2023) 044017}, \href{http://arxiv.org/abs/2305.09172}{{\ttfamily arXiv:2305.09172 [gr-qc]}}.

\bibitem{Sardeshpande:2024bnk}
S.~Sardeshpande and A.~Daripa, ``{A thermodynamic study of ${\textbf {(2+1)}}$-dimensional analytic charged hairy black holes with Born\textendash{}Infeld electrodynamics},'' \href{http://dx.doi.org/10.1140/epjc/s10052-024-13144-3}{{\em Eur. Phys. J. C} {\bfseries 84} no.~8, (2024) 792}, \href{http://arxiv.org/abs/2406.08211}{{\ttfamily arXiv:2406.08211 [gr-qc]}}.

\end{thebibliography}\endgroup
\bibliographystyle{utphys}

\end{document}